\documentclass[usenatbib,usedcolum,usegraphicx]{mn2e}
\usepackage{graphics}

\newcommand{\text}{\textrm}
\newcommand{\F}{${\cal F}$}
\newcommand{\dd}{\mbox{d}}
\newcommand{\obs}{\mbox{\scriptsize obs}}
\newcommand{\GC}{GCS}
\newcommand{\parx}{\varpi}
\newcommand{\X}{\mbox{\boldmath $X$}}
\newcommand{\Y}{\mbox{\boldmath $Y$}}

\newcommand{\Prob}{\mbox{prob}}
\newcommand{\feh}{[\textrm{Fe/H}]}
\newcommand{\mage}{\mbox{age}}
\newcommand{\mtage}{\mbox{\tiny age}}
\title[      Ages for field dwarfs: method and application to
                  the age-metallicity relation]{
         Isochrone ages for field dwarfs: method and
        application to the age-metallicity relation
}

\author[Fr\'ed\'eric Pont and Laurent Eyer]{
     Fr\'ed\'eric Pont$^{1}$\thanks{E-mail: frederic.pont@obs.unige.ch}
              and 
        Laurent Eyer$^{1,2}$\\
  $^1$ Observatoire de Gen\`eve, CH-1290 Sauverny, Switzerland\\
  $^2$ Princeton University Observatory, Princeton, NJ 08544, USA
}

\begin{document}
\maketitle
\label{firstpage}

\begin{abstract}
  A new method is presented to compute age estimates from theoretical
  isochrones using temperature, luminosity and metallicity data for
  individual stars. Based on Bayesian probability theory, this method
  avoids the systematic biases affecting simpler strategies, and
  provides reliable estimates of the age probability distribution
  function for late-type dwarfs. Basic assumptions about the a priori
  parameter distribution suitable for the solar neighbourhood are
  combined with the likelihood assigned to the observed data to yield
  the complete posterior age probability. This method is especially
  relevant for G dwarfs in the 3-15 Gyr range of ages, crucial to the
  study of the chemical and dynamical history of the Galaxy.  In many
  cases, it yields markedly different results from the traditional
  approach of reading the derived age from the isochrone nearest to
  the data point. We show that the strongest effect affecting the
  traditional approach is that of strongly favoring computed ages near
  the end-of-main-sequence lifetime. The Bayesian method compensates
  for this potential bias and generally assigns much higher
  probabilities to lower, main-sequence ages, compared to short-lived
  evolved stages.  This has a strong influence on any application to
  galactic studies, especially given the present uncertainties on the
  absolute temperature scale of the stellar evolution models. In
  particular, the known mismatch between the model predictions and the
  observations for moderately metal-poor dwarfs ($-1<[Fe/H]<-0.3$) has
  a dramatic effect on the traditional age determination.
  
  We apply our method to the classic sample of Edvardsson et al.
  (1993), who derived the age-metallicity relation (AMR) of a sample
  of 189 field dwarfs with precisely determined abundances. We show
  how most of the observed scatter in the AMR is caused by the
  interplay between the systematic biases affecting the traditional
  age determination, the colour mismatch with the evolution models,
  and the presence of undetected binaries. Using new parallax,
  temperature and metallicity data, our age determination for the
  Edvardsson et al. sample indicates that the intrinsic dispersion in
  the AMR is at most 0.15 dex and probably lower. In particular, we
  show that old, metal-rich objects ($[Fe/H]\sim 0.0$ dex, $age > 5$
  Gyr) and young, metal-poor objects ($[Fe/H]<-0.5$ dex, $age < 6$
  Gyr) in many observed AMR plots are artifacts caused by a too simple
  treatment of the age determination, and that the Galactic AMR is
  monotonically increasing and rather well-defined.  Incidentally, our
  results tend to restore confidence in the method of age
  determination from chromospheric activity for field dwarfs.

\end{abstract}

\section{Introduction}

\subsubsection{Isochrone ages for field dwarfs -- Bayesian approach}

Theoretical stellar evolution models have proved spectacularly
successful at explaining the position of stars in the colour-magnitude
diagram (CMD), as a function of only three parameters: mass, age and
metallicity. Comparison of the mean sequences of open and globular
clusters with theoretical isochrones forms the basis of age
estimations in astrophysics. Although second-order discrepancies
subsist such as the subdwarf locus, the slope of the red giant branch,
the width of the main sequence or the distance of the Pleiades, the
method has now been improved to the degree of reaching a relative
accuracy better than 10~percent for cluster ages from theoretical
isochrones (see e.g. Rosenberg et al. 2002 for globular clusters).

In principle, this method can also be applied to individual stars. The
model predictions can be interpolated to associate a given star's
observed parameters, by inverting the relation given by the models
between the physical parameters (mass, temperature, abundances) and
observable parameters such as luminosity, temperature, metallicity,
colour and magnitude. In particular, because late-F and G dwarfs have
lifetimes comparable to the age of the Galaxy, deriving individual
ages for field F and G dwarfs in the solar neighbourhood is of crucial
importance to study the chemical and dynamical history of the Galactic
disc. In practice, however, deriving ages for late-type dwarfs turns
out to be difficult. For the ages typical of galactic populations,
from $\sim$ 1 Gyr up to the maximum age of stars in the Galaxy, the
model isochrones are separated only by small distances in observable
parameter space on and near the main sequence, and high accuracy on
the temperature, distance and metallicity determinations, as well as
high confidence in the absolute temperature and metallicity scales of
the models and observations, are needed to obtain meaningful results
for individual ages.

One landmark study in that field is Edvardsson et al. (1993,
hereafter {\bf E93)}, who obtained very accurate multi-element
abundances for 189 field F and G stars and computed ages spanning the
whole lifetime of the Galactic disc. They derived the ages for
individual stars in their sample by comparison with Vandenberg (1985)
isochrones. Their results have subsequently been recomputed with more
recent Bertelli et al. (1994) isochrones by Ng \& Bertelli (1998).
Chen et al. (2000) have added multi-element metallicity data
for 90 more disc stars, and Bensby, Felzing \& Lundstr\"om (2003) for 66 stars.

Following the availability of Hipparcos parallaxes for most nearby F
and G dwarfs, ages have been derived for much larger sets of data
(e.g.  Asiain et al. 1999; Feltzing et al. 2001; Ibukiyama \& Arimoto
2002). The publication of the large Geneva-Copenhagen
solar-neighbourhood survey (Nordstr\"om et al. 1999, hereafter \GC),
with metallicities, distances and Str\"omgren photometry for more
than 16'000 local F and G dwarfs, is likely to prompt more such
studies in the near future.

Another important method to derive ages for field F-G dwarfs is the
use of chromospheric activity as an age indicator (e.g. Kraft 1967,
Noyes et al. 1984). It has been applied to a large sample of disc F
and G dwarfs by Rocha-Pinto et al. (2000).  The results of the
isochrone and chromospheric age estimates are not showing satisfactory
agreement.



Deriving age estimates from isochrones for individual stars is an
inverse problem. The tracks calculated from theoretical evolution
models define a function $\Y={\cal F}(\X)$ relating the input physical
parameters $\X$ (age, mass, abundance) to the observable parameters
$\Y$ (e.g. temperature, luminosity, metallicity). The objective is to
derive estimates of the physical parameters $\X$ from the observed
$\Y$.

All studies quoted above have estimated isochrone ages for individual
stars by selecting the isochrone nearest to the object in data space,
i.e. computing ${\cal F}^{-1} (\Y)$. The uncertainties affecting the resulting ages are estimated
by probing the ${\cal F}^{-1}$ around ${\cal F}^{-1}(\Y)$ according to
the uncertainties affecting $\Y$.

However, in practice, there are two conditions for the inverse
function ${\cal F}^{-1}$ to provide an unbiased estimator of the real
$\X$:
\begin{enumerate}
\item the function must be reasonably linear within the uncertainties on
    $\Y$.\\
\item the uncertainties on $\Y$ must be much smaller than the
    variations in its a priori probability distribution.
\end{enumerate}

These two conditions are generally satisfied for the age determination
of early-type stars at the bright end of the main sequence, with ages
in the range 0-3 Gyr.  In this regime, the observational uncertainties
are generally smaller than the range over which the isochrones curve,
and over which the a priori density varies. It is much less clear that
the conditions are respected for later-type dwarfs, with ages in the
3-15 Gyr range. In this range, condition (1) is not satisfied, because
the isochrones are densely packed and highly curved within the
observational uncertainties. Condition (2) is not satisfied either:
under any reasonable assumption, the a priori distribution of the
observables can vary enormously within the uncertainties on $\Y$. For
instance, varying the luminosity within one or two sigma can move an
observation from the main-sequence into the Hertzsprung gap, where the
a priori density is an order of magnitude lower, or even below the
ZAMS, where the a priori density is zero. Therefore, the isochrone age
from the simple inversion of the function defined by the evolution
models can suffer from significant systematic biases. Because those
objects are the most interesting to study the history and evolution of
the Galaxy (age range 3-15 Gyr), it is important to derive realistic
and unbiased age estimates for them.

When condition (1) is not satisfied, one way to obtain an unbiased
estimator is to consider the complete probability distribution
function ('pdf') of $\Y$ given its uncertainties, $\Prob (\Y)$, and to
compute the corresponding distribution of $\Prob \left( {\cal
    F}^{-1}(\Y)\right)$. The resulting probability distribution
function is known as the likelihood.  Methods based on the likelihood
can provide reliable estimators when the function relating the
observed values to the unknown parameters is highly non-linear.

When condition (2) is not satisfied, even estimators based on the
likelihood will be biased. One common way to deal with the problem is
to build simulations of the whole procedure, estimate the systematic
biases from the simulations, and then correct the results with a
posteriori compensations for the biases. However, Bayesian probability
theory offers a much more robust method to obtain unbiased estimates
in that case. The Bayesian approach includes both the likelihood and
the a priori distribution of the parameters to compute the complete,
unbiased posterior probability distribution.

When possible, fully Bayesian analyses are often avoided because of
their large demand in computing time and their conceptual difficulty.
In recent years however, their use has become more and more frequent,
following both the increase in computing power and the development of
the theory. For a general introduction to Bayesian data analysis see
Sivia (1996), for a very clear and detailed presentation see Jaynes
(2003), and for a rapid overview centered on astrophysical
applications see Loredo (1990). In this paper, we assume that the
notations and concepts of probability theory are familiar to the
reader. We are mostly using notations as in Sivia (1996).  Cases that
necessitate a Bayesian analysis are rare. In general, measurement
uncertainties are much smaller than the range of the a priori
distribution. Bayesian analyses are necessary in the case of high
relative errors on single-case estimates. They have been used to
derive the most likely value of the cosmological parameters (e.g.
Slosar et al.  2003) and to analyse the neutrino data of SN1987A
(Loredo 1990).  Another example is the case of the study of the bias
affecting trigonometric parallaxes (Lutz \& Kelker 1973): In the wake
of the {\it Hipparcos} astrometric satellite mission, an extensive
literature has appeared on the subject (Oudmaijer, Groenewegen \&
Schrijver 1998; Pont 1999; Reid 1999; Arenou \& Luri 2002; Smith 2003). 
In that case, it was found that with
high-relative-error parallax measurements ($\sigma_\parx/\parx > \sim
0.2$), straightforward statistical methods brake down, and the effect
of the prior begins to become so dominant that hardly anything
meaningful can be derived about the value of the distance
independently of the a priori assumptions. It is now accepted that a
limit of $\sim$10 percent has to be put on $\sigma_\parx/\parx$ to
derive robust distance estimates from trigonometric parallaxes
independently of further assumptions. The Lutz-Kelker effect is a
typical Bayesian effect showing when the shape of the prior has to be
taken into account.

In this paper, which is intended both as a reconsideration of previous
data and as a preparation for the coming analysis of new large
samples, we point out fundamental features in the analysis of ages for
individual stars that have been overlooked by all previous studies
that we are aware of, and that can profoundly affect the results. We
examine in what way the age determination can be improved using
probability theory.  We propose a method to compute the Bayesian age
probability distribution for field stars and compare them to
likelihood estimates (Part One). We then focus on the E93 sample as an
illustrative application (Part Two).

\subsubsection{The E93 study and the Galactic age-metallicity relation}
\label{introe93}

The E93 sample of 189 stars in the solar neighbourhood still provides one
of the most accurate and extensive bases for the study of the chemical
evolution of the Galactic disc. In particular, many authors
subsequently accepted their age-metallicity plot (repeated in
Fig.~\ref{amr_e93}a, but see companion figures) and their
interpretation of a very substantial spread of $\feh$ for a given age
in the solar neighbourhood. The plot shows a large intrinsic scatter
at ages between 3 and 10 Gyr, with a standard dispersion of $\simeq
0.24$ dex and almost no age-metallicity relation in this range. The
age-metallicity plot for their sample is often presented as "the
age-metallicity relation of the Galaxy" and a dispersion of the order
of 0.25 dex -- corresponding to a total range of $\sim$ 0.6 to 0.8 dex
-- in metallicity at a given age has been taken as an observational
requirement that the models of the Galaxy are requested to meet (e.g.
Carraro et al. 1998). However, more recently, Garnett \& Kobulnicky
(2000) have revealed an important dependence of the metallicity
dispersion with distance in the E93 sample, indicating that the age
metallicity relation (AMR) from E93 may have been affected by strong
systematic biases.

Rocha-Pinto et al. (2000) have studied the dispersion of the AMR with
age determinations based on chromospheric activity. Their ages are
only weakly correlated with the isochrone ages, and they find a low
intrinsic dispersion of the AMR. This result is particularly
significant given the statistical implausibility of {\it decreasing}
the dispersion of the AMR with added uncertainties.

The age-metallicity relation is one of the most important
observational constraints on models of the evolution of the Galaxy. It
expresses how stellar formation has enriched the interstellar medium
over time, and therefore depends on the star formation rate, the
chemical yields, the efficiency of recycling, infall and outflow, and
the amount of mixing in the gas. Because the shear induced by
differential galactic rotation spreads stars and gas around all
galactic azimuths in a few rotations, theoretical models commonly
assume a metallicity depending only on time and galactocentric radius
(e.g. Chiappini et al. 1997). Several observations support this
assumption. The inhomogeneities in the abundance of the interstellar
medium are small (Kobulnicky \& Skillman 1996; Meyer, Jura \& Cardelli
1998). $R\sim R_0$ Cepheids show a low abundance dispersion
(Andrievsky et al. 2002); young open clusters show an abundance
dispersion lower than 0.20 dex (Twarog 1980; Carraro et al.  1998).
The indication from older open clusters is more ambiguous (Twarog 1980;
Piatti et al. 1995; Carraro et al. 1998; Friel et al. 2002), showing a
large scatter that can be mostly attributed to a radial metallicity
gradient. Kotoneva et al. 2002a find that only a modest intrinsic
scatter in the AMR is needed to fit the solar neighbourhood K-dwarfs
data. In external spiral galaxies, the abundance dispersion of young
features at a given galactocentric distance is typically much smaller
than 0.2 dex (e.g. Kennicutt \& Garnett 1996 for HII regions in M101).

In this context, the E93 result comes as a surprise, and a very
difficult requirement to be met by the models. E93 computed the
present Galactic orbits of their objects, and concluded that only a
small part of the observed metallicity dispersion was due to orbital
diffusion i.e. the fact that metal-richer and metal-poorer stars born
at lower or higher galactocentric radii cross the solar neighbourhood.
The remaining dispersion, covering a total range of the order of
0.6-0.8 dex at a given age, was taken to be the indication of a very
high dispersion of the metallicity of the gas at a given time and
galactocentric radius (or even a complete lack of correlation between
age and metallicity, e.g. Feltzing et al. 2001). Such a result implies
a very inefficient azimuthal mixing of the gas in the disc, or
extremely frequent infall episodes, with pockets of gas of very
different metallicities sharing a similar radius at a given time.

An alternative explanation for a high intrinsic dispersion was
explored by Selwood \& Binney (2002) and L\'epine et al. (2003), who
showed that radial migration of the stellar orbits without
conservation of the angular momentum, for instance under the influence
of spiral arm perturbations, could move stars in galactocentric radius
by several kiloparsecs within the Galactic disc. This could bring
together stars of the same age on similar orbits but formed at very
different galactocentric radii, therefore with very different
metallicities because of the radial metallicity gradient in the disc
of the Galaxy.
 
In the second part of this article, we apply our age determination to
the E93 sample as an illustration of our approach to the age
determination, and conclude that -- as correctly anticipated by
Garnett \& Kobulnicky (2000) -- most of the "cosmic" dispersion in the
AMR of E93 is due to uncertainties in the ages, and that the data
actually indicate a cosmic dispersion of less than 0.15 dex.

\section*{PART ONE: Age estimates from isochrones for individual late-type stars}

\section{Theoretical basis}
\label{theory}
\subsection{Confrontation of the standard and Bayesian approach}

\subsubsection{The standard approach}
Stellar evolution models define a function relating physical
parameters to observable quantities:
\[
\Y = {\cal F} (\X)
\]
where $\X$ are the physical input parameters, namely mass, age and
abundance: $\X\equiv (m,t,z)$, and $\Y$ are the observed quantities --
observed or inferred directly from observations -- e.g. temperature,
luminosity and metallicity: $\Y\equiv (T, L, \feh)$.

The standard approach to computing the age of an individual star from
theoretical isochrones is to interpolate the stellar evolution tracks
to find which age and mass value correspond to the same point as the
star in the $(T, L, \feh)$ space.

Interpolation between the models is needed to yield a value of \F\ for
all $(m,t,z)$ triplets\footnote{In this section we assume that for a
  given object, values of $T_{\obs}$, $L_{\obs}$ and $\feh_{\obs}$ are
  obtained from the observations (the transformations from observed
  colours, magnitudes and parallax to $T_{\obs}$ and $L_{\obs}$ are
  considered reliable). The reasoning would be exactly the same if we
  use colour and magnitude instead of temperature and luminosity.}.
Given the observed values $T_{\obs}$, $L_{\obs}$ and $\feh_{\obs}$,
the standard approach thus inverts the relation \F\ to find
\[
(m_o,t_o, z_o) = {\cal F}^{-1}(T_{\obs}, L_{\obs}, \feh_{\obs})
\]
where the "o" subscript denote the values for the considered object.

The function \F\ is not strictly bijective because isochrones do
sometimes cross each other in the $\Y$ space.  ${\cal F}^{-1}$ can be
uniquely defined nevertheless by considering, when it is multiply
defined, only the stage that is more slowly evolving. ${\cal F}^{-1}$,
of course, is also undefined in the large portion of the $\Y$ space
that do not contain any evolution tracks.

A simple way to estimate the uncertainties on the $t_o$ age obtained
by the inversion of the \F\ function is to calculate the value on
${\cal F}^{-1}$ found by moving the data point according to the
observational errors:
$$
(m_{\pm \sigma},t_{\pm \sigma},z_{\pm \sigma})=  
                        {\cal F}^{-1}(T_{\obs} \pm \sigma_{\tiny T},
                                      L_{\obs}\pm \sigma_{\tiny L},
                                      \feh_{\obs}\pm \sigma_{\tiny [Fe/H]})
$$
either one at a time or all at the same time. 

A slightly more sophisticated approach is to compute ${\cal F}^{-1}$
over the whole $(\log T, L, \feh)$ space, and to assign to each point
the probability given by the distribution function of the
observational uncertainties. For instance, if the observational
uncertainties are described by Gaussian functions with dispersions
$\sigma_{\tiny [Fe/H]}$, $\sigma_{\tiny \log T}$ and $\sigma_{\tiny
  \log L}$, then the recovered age distribution function is based on
the {\it likelihood} function:
\[
{\cal L}(T,L,\feh)=\ 
\frac{1}{\sigma_{\tiny [Fe/H]}\,\sigma_{\tiny T}\,\sigma_{\tiny L}\,(2\pi)^{3/2}}
\]\[
\times \exp\frac{-(\log T_{\obs}-\log T)^2}{2 \sigma^2_{\tiny \log T}}
\exp\frac{-(\log L_{\obs}-\log L)^2}{2 \sigma^2_{\tiny \log L}} 
\] \begin{equation}
\times \exp\frac{-(\feh_{\obs}-\feh)^2}{2 \sigma^2_{[Fe/H]}}\
\label{likelihood}
\end{equation}

This likelihood is the conditional probability of a point being
observed at $(T_{obs},L_{obs},\feh_{obs})$ given a true value of
$(T, L, \feh)$, or $ {\cal
  L}(\textrm{observed,true})\equiv \Prob(\textrm{observed} |
\textrm{true}) $ where the "$\mid$" symbol denotes conditional
probabilities.  The terms on the right result from the Gaussian
distribution of the uncertainties.  Instead of simply inverting the
\F\ function at the value of the data point, an age pdf can be
obtained from the histogram in age of the likelihood over all possible
ages:
\[
{\cal L}_t(t) = 
\int_{R} {\cal L}(T, L, \feh )\, \dd \feh\, \dd T\, \dd L
\]
where $R$ is the region in ($T, L, \feh$) space where the ${\cal
  F}^{-1} (T,L,\feh)=t$. The maximum of ${\cal L}_t$ can be used as an
estimator (maximum likelihood method).

\subsubsection{Bayes' theorem}
\label{bayesth}

However, $\Prob(\textrm{observed} \mid \textrm{true})$ is not really
the probability that one is trying to determine when performing a
measurement. One is not attempting to estimate the probability of the
observed value assuming the true value, but indeed the value of the
true quantity, given the observation, i.e. $\Prob(\textrm{true} |
\textrm{observed})$.

These two quantities are related through Bayes' theorem:
\begin{equation}
\Prob(H|D)=\frac{\Prob(D|H)}{\Prob(D)}\, \Prob(H)
\label{bayes}
\end{equation}
where $H$ can be any set of hypotheses (in our case the true age
value) and $D$ the observed data. The term on the left is called the
posterior probability (the probability of $H$ given $D$, which is what
one wants to determine), the numerator on the right is the likelihood
(the probability distribution of $D$ assuming $H$, or the "likelihood"
of observing $D$ if $H$ is true). $\Prob(H)$ is the prior probability
(what was known about $H$ a priori), and $\Prob(D)$ is a normalizing
factor independent of $H$ (that can be ignored for our purposes).
Therefore, according to Bayes' theorem,  the condition for the
  likelihood to be a good estimator of the posterior pdf is that the
  prior pdf can be neglected.  Then Bayes'
theorem becomes $\Prob (H|D)\sim \Prob(D|H)={\cal L}(D,H)$.
  
The basic criterion to determine whether the prior probability
distribution can be neglected in a given problem is to compare the
scale of variation of the prior pdf with the scale of the
observational uncertainties. If the uncertainties are much smaller
than the scale over which the prior varies, then the likelihood
"overwhelms" the prior. This is the case for instance when the
magnitude of a star is measured with an accuracy of, say, 0.01 mag.
The prior pdf can vary from case to case, but a natural prior is to
assume nothing on the magnitude, allowing for the star to be located
at random in a 3-D space, which implies for the prior a dependence in
$10^{0.6 \cdot \Delta m_v}$, a variation of a factor $\sim$1.4 percent
for each 0.01 magnitude. Thus the variation of the prior is negligible
compared to the variation of the likelihood ($\sim$ 30 percent on
$\sigma=0.01$ mag for a Gaussian distribution).
  
In some cases though, the prior pdf cannot be neglected. The prior
probability varies a lot over the span of the observational
uncertainties. In other words, previous knowledge indicates that one
part of the likelihood distribution is much less probable than the
other.  For example, imagine measuring with very low accuracy the
magnitude of a star picked up at random from the HD catalogue, and
obtaining $m_V=7.0\pm1.0$ mag.  The likelihood is a Gaussian,
$N(7,1)$. It would not result however that $6.0$ and $8.0$ are equally
likely for the star's actual magnitude, given the fact that faint
stars are more numerous than bright stars by a factor $10^{0.6 \cdot
  \Delta m_v}$ in a magnitude-limited survey (that is the prior
$\Prob(H)$). After the measurement, $m_V>7$ is a much better estimate
of the true magnitude, $\Prob(H|D)$, than $m_V<7$. At the other
extreme, values $m_V\sim 9$ are much less likely than indicated by the
function $N(7,1)$, because the HD catalogue has a magnitude limit
$m_V\sim 8.5$. In this example the likelihood was not peaked enough to
"overwhelm" the prior, and the posterior pdf obtained through Bayes'
theorem resembles the prior more than the likelihood.

Another, more astrophysically relevant, such situation is the case of
trigonometric parallaxes with high relative errors (see Introduction
for references). The prior distribution of parallaxes for a single
object is very steep: knowing that parallaxes are the inverse of
distances and that space is 3-dimensional, it can be inferred that
lower values of the parallax $\parx$ are more likely a priori by a
factor $\parx^{-4}$ for a given object. Only if the parallax error is
much smaller than the range over which this factor changes
($\sigma_\parx/\parx<\sim 10$ percent) does the prior knowledge become
irrelevant, and one can proceed to derive an unbiased estimate of the
distance.

When the conditions for neglecting the prior are not satisfied, one
way to proceed is to use likelihood estimators anyway, and to deal
with the biases introduced by neglecting the prior with methods like
weighted statistical indicators, ad hoc empirical corrections, or bias
corrections from Monte Carlo simulations. This is often the only
alternative when the mathematical structure of the problem is complex
and there is no clear way to characterize the prior $\Prob(H)$.

However, equation~(\ref{bayes}) provides the means to calculate the
correct unbiased posterior pdf if a functional form of the prior can
be given. The prior is in general not known exactly, but that is often
not necessary. A reasonable approximation is generally
sufficient\footnote{If it is not, it means that the result will be
  more sensitive to the prior that to the likelihood. In plain
  English, more sensitive to what was known a priori than to the
  measurement. In that case (to paraphrase Loredo 1990), maybe one
  should consider getting better measurements!}. In the case of
parallaxes, assuming that space is 3-dimensional and Euclidean is
enough to give a good prior for the analysis of the Lutz-Kelker bias.
In the case of the ages, reasonable assumptions on the prior can be
made, for instance a flat age distribution and some power-law mass
distribution.

It is important to remember that although the dependence of the posterior
pdf on the prior pdf could be taken as a drawback of the Bayesian
approach, the likelihood estimates are independent of the prior only
in appearance. In reality, they make a much more obviously invalid
-hidden- assumption about the prior by implicitly assuming a flat
prior {\it in the space of the observable data}.  In the case of the
ages, that means assuming that the HR-diagram is uniformly filled,
which is obviously very far from true.  This hidden assumption is of
no consequence in cases when the experimental errors are very small
compared to the changes in density in the HR-diagram. However, for
late-type dwarfs, that is not the case. The whole width of the main
sequence is only a few times the size of the observational errors, and
the a priori density varies a lot within the error intervals. In that
case even a low-information prior like a flat age distribution is much
better than a flat prior distribution in data space implicit in
likelihood methods.

\subsubsection{Bayesian age estimates from isochrones}

\begin{figure}
\caption{Representative example of a colour-magnitude diagram for
  dwarfs in the solar neighbourhood (from \GC) with metallicities near
  solar. The isochrones from Girardi et al. (2000) for z=0.02 are
  overlaid.  Typical observational uncertainties are illustrated for
  one data point in the post-main sequence zone. The age probability
  distributions for this point are given in the lower panel. Dashed line:
  Gaussian distribution with 10$\pm 2$ Gyr. Dotted line: likelihood pdf
  for the sample point. Solid line: posterior age pdf.}
\resizebox{\hsize}{!}{\includegraphics{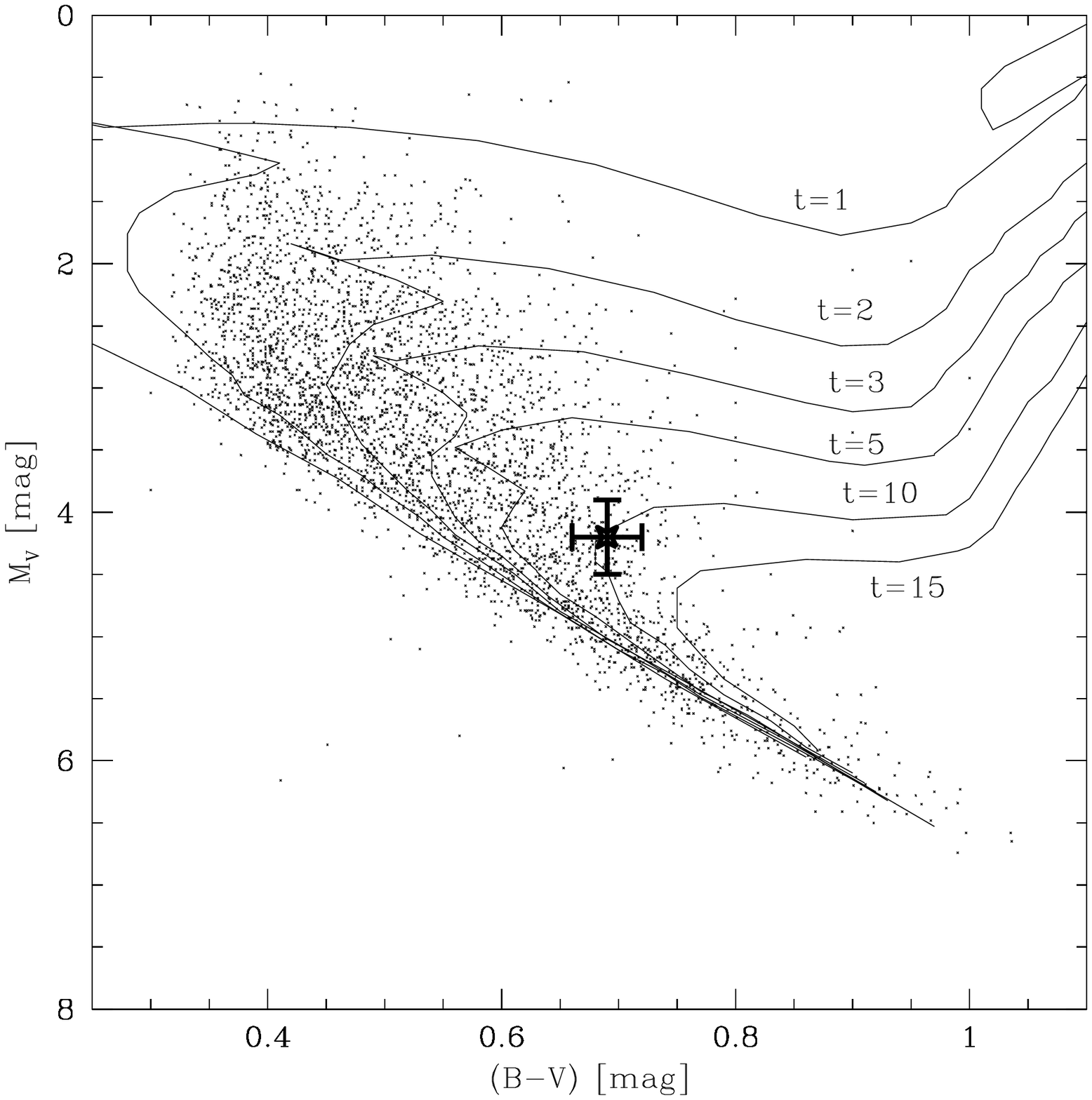}}
\resizebox{\hsize}{!}{\includegraphics{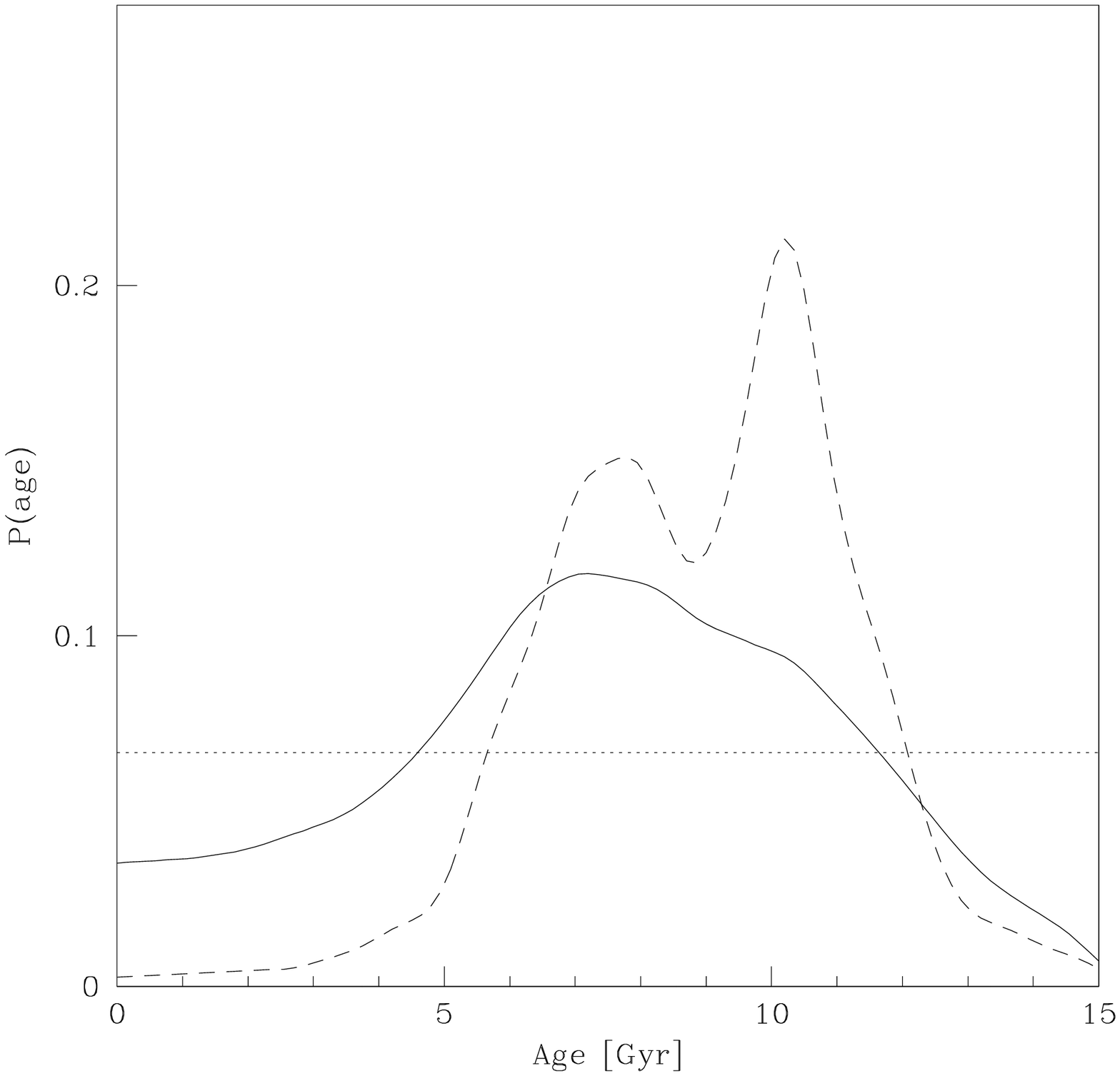}}
\label{illus}
\end{figure}

Fig.~\ref{illus} illustrates a representative example for the
solar-neighbourhood: deriving ages from the observed colour and
magnitude is done by comparison of the data with isochrones from
theoretical evolution tracks. Let us ignore the metallicity dimension
for the time being. The background dots in Fig.~\ref{illus} display a
typical distribution of stars in the HR-diagram for a
magnitude-limited sample of the Galactic disc (from \GC), with a dense
main-sequence and sparsely populated Hertzsprung gap, and a
superposition of many "turnoffs" due to the mixture of stars with a
wide range of ages. The error bars on the measured parameter are shown
for a single point. This point is located on the 10 Gyr isochrone and
therefore has ${\cal F}^{-1} (T, L)$=10 Gyr.

Focusing on this data point, we first note that the observational
errors are large compared to the regions where the \F\ function can be
linearised. This implies that the description of the age probability
distribution by a central value and a single error bar obtained by the
propagation of the uncertainties on the observed parameters $T$ and
$L$ will not be a good representation. In particular, the age pdf can
be expected to have a wide tail towards lower age values. Although the
one-sigma interval remains within $t=10\pm$2 Gyr, the 3-sigma interval
reaches $t=0$ Gyr.  This asymmetry can be taken into account by
computing the complete age likelihood pdf ${\cal L}(t)\equiv
P(T_{obs},L_{obs}|\, t)$, plotted as a dotted line in the lower panel
of Fig.~\ref{illus}.

However, as was reminded in the previous section, the likelihood can
still be a biased estimator of the actual probability distribution of
the real age if the uncertainties do not make the influence of the
prior pdf negligible. If our point is a random representative of the
sample it was picked up from, then the prior pdf resembles the density
of background points in the $(T,L)$ plane (see Section~\ref{prior} on
more details on the computation of the prior). The prior is seen to
vary greatly within the span of two times the observational errors or
less, dropping by a large factor when moving from the main-sequence
zone to the subgiant zone (Hertzsprung gap). Clearly, the conditions
for using the likelihood as an estimator are not satisfied and any
estimator based on the likelihood only will be biased.

The fundamental reason for this is the acceleration of stellar
evolution after the main sequence phase. A star spends much more time
on the slowly-evolving part of the main sequence than on the rapidly
evolving subgiant zone. Therefore, a mixed-age population will be much
more dense on the main-sequence than above it. As a consequence any
point observed above the main sequence, with error bars that are of
the order of the distance separating it from the slow-evolving zone,
has considerable probabilities of being actually located on the main
sequence and brought where it is observed by observational errors
("contamination"). The likelihood does not take this into account and
gives equivalent probabilities to positive and negative errors. The
prior term $\Prob(H)$ of Bayes' Theorem is what allows this to be
taken into account.

The lower panel of Fig.~\ref{illus} compares the Bayesian posterior
pdf $\Prob(t|T_{obs},L_{obs})$ for our sample point to the
distribution of the likelihood.  The age probability is significantly
shifted towards the slow-evolving main sequence. Values between 0 and
5 Gyr, that were practically excluded by the ${\cal F}^{-1}$ and
likelihood estimators, now have significant probabilities, and the
median of the pdf is shifted from 10 to 7.5 Gyr, implying a systematic
bias of 2.5 Gyr or 25 percent.

In this 2-D illustration we did not take the metallicity into account.
Note however that uncertainties on the metallicity are also of the
same order of magnitude as the expected variations of the metallicity
prior pdf, so that likelihood-based estimates will also suffer from
metallicity-dependent biases (see Section~\ref{prior}). In the next
section, we consider the issue analytically in the even more
simplified 1-D case, before moving to a more complete calculation in
Section~\ref{compute}.

\subsection{Simplified 1-D approach}
\label{laurent}

By reducing the dimensionality of the problem, we can illustrate some
fundamental statistical features of the age determination. Les us assume
here that the age, $t$, is computed from a
single observable parameter,  the logarithm of luminosity
$\log L$.  The objective is to estimate the probability distribution
of the real age, $t$, given the observed value $\log L_{\obs}$ and
the transformation law $\log L={\cal F}(t)$. This is done through Bayes'
theorem:
%
$$
 \Prob\left(   t \mid  \log L_{\obs} \right)
 = \frac{  \mbox{prob}(\log L_{\obs} \mid t)  }{
           \mbox{prob}(\log L_{\obs})         }
 \mbox{prob}(t)
 $$
Without prior knowledge on the real age, we assume that the prior
probability $\Prob(t)$ has a flat distribution between $0$ and
$t_{\mbox{\scriptsize max}}$. The difficulty is that the likelihood is
expressed in the parameters of the observables, while the prior is
expressed in terms of the physical parameter $t$. Since Bayes' theorem
has to be expressed in a coherent set of variables, the probability
distribution functions have to be modified accordingly (using the
transformation \F). In one dimension, this is done by the chain
rule:
$$
\mbox{prob}(\log L) \, \dd\log L = \mbox{prob}(\log L) \mid
                          \frac{\dd{\cal F}(t)}{\dd t}\mid \dd t
$$
where $\mbox{prob}(\log L)$ should be expressed in the variable $t$.

\begin{figure}

  \resizebox{\hsize}{!}{\includegraphics{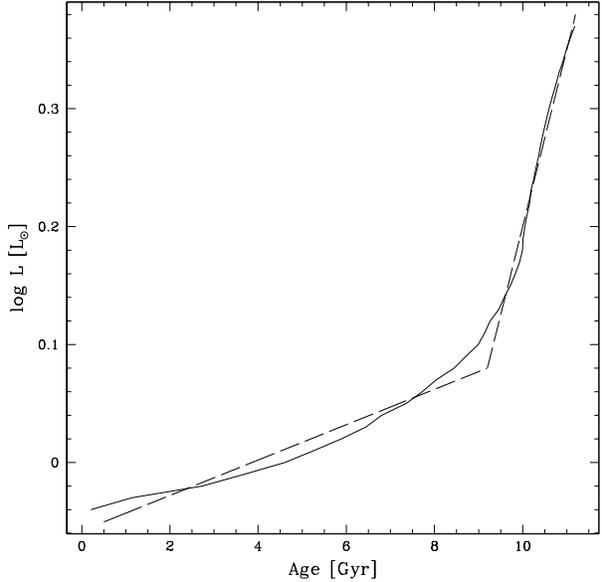}}

   \caption{Solid line: Transformation law \F, $\log L={\cal F}(t)$,
     i.e. the logarithm of the luminosity, $\log L$, as function of
     the age, $t$, at fixed solar temperature (5780 K) from Geneva
     stellar evolution models (Schaller et al. 1992).  Dashed line:
     simple modelisation of ${\cal F}$ in two regimes.  The slope of
     ${\cal F}$, low below 9.2 Gyr (main-sequence phase), becomes much
     higher above 9.2 Gyr (evolved phase).  The ratio between the
     speed of evolution (ratio of the slopes) is about a factor 10.}
  \label{fig:tlogl}

\end{figure}

An actual example of the transformation between $t$ and $\log L$ is
shown in Fig.~\ref{fig:tlogl}. The \F\ relation at solar temperature
and metallicity can be approximated by two linear relations with a strong
change of slope at the end of the main sequence:

$$\log L={\cal F}(t)=\left\{ \begin{array}{ll}
                      \alpha t + \gamma           & \mbox{if $0 < t \leq t_0$} \\ 
                      \beta (t-t_0) + \alpha t_0 + \gamma & \mbox{if $t_0 < t < t_{\mbox{\tiny max}}$} 
                   \end{array} 
            \right.
$$
with $\beta \gg \alpha$. In our example, $\alpha = 0.015$,
$\beta=0.15$, $\gamma=-0.058$, $t_0= 9.2$~Gyr and $t_{\mbox{\tiny
    max}}=11$~Gyr.

We further assume that the noise measurement on $\log L$ has a
Gaussian probability distribution with standard deviation
$\sigma_{\log L}$.

 In this case the likelihood is expressed as:
$$
\mbox{prob}(\log L_{\mbox{\tiny obs}} \mid \log L) =
   \frac{1}{ \sigma_{\tiny \log L}\, \sqrt{2 \pi}} \exp -\frac{(\log L_{\mbox{\tiny obs}}-
     \log L)^2}{2\sigma_{\tiny \log L}^2}
   $$
The uniform prior probability distribution of the
age $t$ is transformed, using the chain rule, into the prior probability
distribution expressed in $\log L$:

$$
\mbox{prob}(\log L) = \left\{ \begin{array}{ll}
					c_1    	 & \mbox{if $\log L \leq \log L_0$} \\
				        c_2      & \mbox{if $\log L > \log L_0$} 
			        \end{array}
	                 \right.
$$
With $c_1 = (\beta/\alpha) c2$ and $\log L_0 ={\cal F}(t_0)$.
The prior expressed in $\log L$ is therefore a step function.

\begin{figure}

  \resizebox{\hsize}{!}{\includegraphics{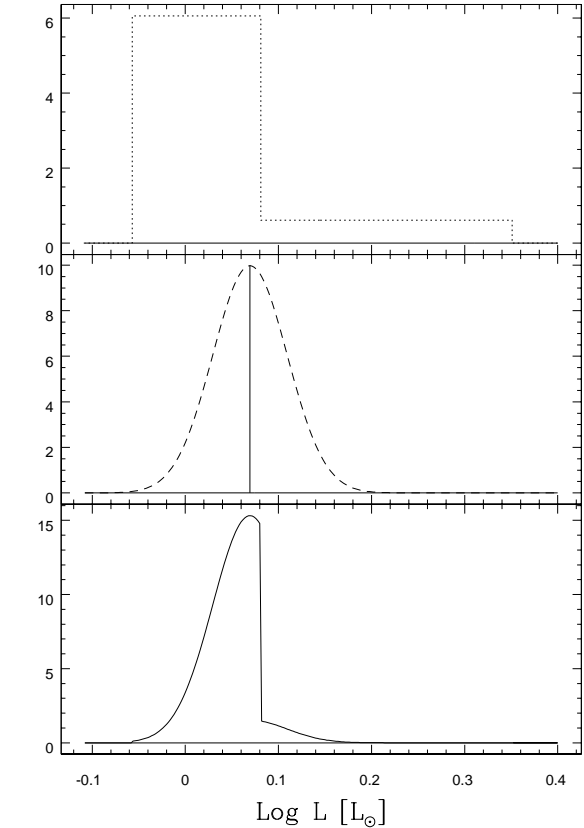}}
  \resizebox{\hsize}{!}{\includegraphics[angle=0]{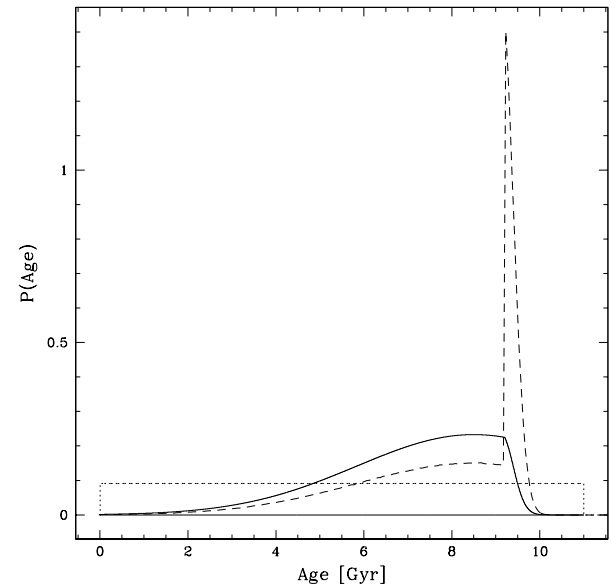}}
   \caption{Upper panel: 
            Prior of uniform age (dotted line),
	    likelihood (dashed) and posterior (solid line) probability distribution functions,
            all expressed in $\log L$.
	    Lower panel: Prior (dotted line), likelihood (dashed line) and posterior
	    probability distribution (solid line) functions expressed in age.}
  \label{fig:demarche}

\end{figure}

Finally, the posterior pdf can be expressed in the age variable, $t$: 


%


$$
\begin{array}{l}    
   \mbox{prob}(t|t_{\mbox{\tiny obs}}) = \\
   \\
    \left\{ \begin{array}{ll}
	\sim  \frac{\alpha c1}{ \sigma \sqrt{2 \pi}}
          \exp -\frac{(\alpha t +  \gamma -\log L_{\obs})^2}{2 \sigma^2}
                        &  \mbox{if $0 < t \leq t_0$} \\
        \sim  \frac{\beta c2}{ \sigma \sqrt{2 \pi}}
          \exp -\frac{(\beta (t-t_0) + \alpha t_0 + \gamma -\log L_{\mbox{\tiny obs}})^2}{2 \sigma^2}
                        &  \mbox{if $ t_0 < t < t_{\mbox{\tiny max}} $} \\
			0
			& \mbox{otherwise} 
            \end{array} \right.
\end{array}
$$ 
where $\sigma=\sigma_{\log L_{\obs}}$.

Let us consider a measurement obtained at the value $\log
L_{obs}$=${\cal F}(t\!=\! 8.5$ Gyr).  Fig.~\ref{fig:demarche}
shows the corresponding prior, likelihood and posterior probability
distributions in terms of $\log L$ and of $t$.

Several effects are apparent:
\begin{enumerate}
\item The transformation laws ${\cal F}^{-1} (\log L)$ and ${\cal F
    }(t)$ can qualitatively modify the probability distributions. Even
  if the likelihood pdf has a Gaussian shape in the variable $\log L$,
  it can be drastically different when expressed in age.
\item In the literature, the quoted $\sigma_{-}, \sigma_{+}$ around
  the estimated age are often the simple transformation of the two
  values $mean_{\log L}-\sigma_{\log L}$ and $mean_{\log
    L}+\sigma_{\log L}$ of the likelihood function through ${\cal
    F}^{-1}$. These are {\it not} necessarily directly related to the
  quantiles or standard deviation of the posterior age pdf. The
  posterior pdf may have a non-Gaussian shape, its mean value, moments
  and quantiles may be all modified by the prior and by the
  transformation law \F. In the example of Fig.~\ref{fig:demarche} the
  posterior pdf has become very asymmetric.
\item The prior pdf expressed in the variable $\log L$ is unevenly
  distributed, with a lower probability for $\log L$ for ages higher
  than 9.2 Gyr. This results in a strong decrease of the posterior pdf
  compared to the likelihood for ages higher than 9.2 (see below).
\end{enumerate}

\begin{figure}

  \resizebox{\hsize}{!}{\includegraphics{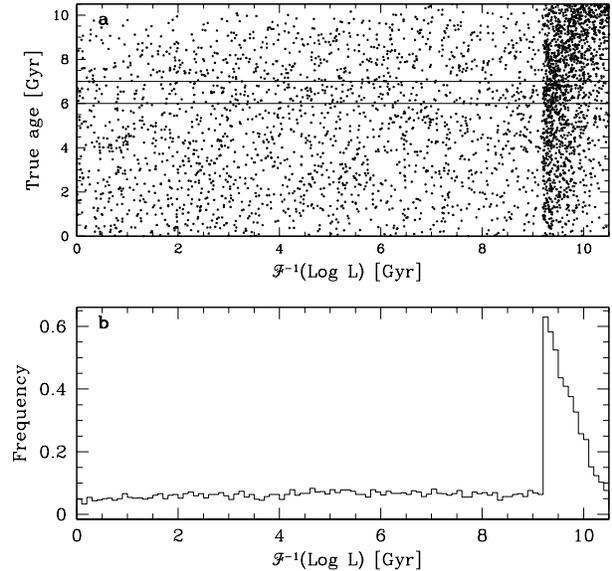}}
   \caption{(a) 
     Uniform random sample of true ages, as a function of the age
     determined with the direct method \F$^{-1}(\log L)$.  The
     discontinuity of the distribution of ages is clearly visible at
     the age corresponding to the end of the main sequence.  We call
     this phenomenon the terminal age bias. (b) Histogram of ages from
     the direct method with true ages between 6 and 7 Gyr,
     as indicated in (a).}
  \label{fig:termage}

\end{figure}

In this 1-D model, ages that are not calculated though Bayes' theorem
are subject to a systematic bias that we call the "terminal age bias".
"Terminal age" refers to the age for which the \F\ relation changes
slope, roughly corresponding to the end-of-main-sequence lifetime.
Fig.~\ref{fig:termage} illustrates this bias by plotting the age
${\cal F}^{-1} (\log L)$ against the real age for a randomly drawn
sample with $0<t<t_{max}$. The histogram of the likelihood ages is
given for the real age interval $6<t<7$, showing the strong excess of
the likelihood ages near the terminal age.

The biased nature of the likelihood-based method is apparent. Maximum
likelihood gives the "best" solution in the sense of taking the most
probable value of the likelihood function. However, it does not take
into account the fact that some values of $\log L$ are more likely a
priori due to the shape of the \F\ function, in our example that
contamination from the slower-evolving main-sequence is important in
the subgiant zone.

When using a maximum likelihood method, systematic biases have to be
evaluated with additional Monte-Carlo simulations, and the estimator
corrected if necessary. Additional knowledge of the system that was
not used to compute the statistical estimator is added in the
Monte-Carlo simulations.

Bayesian methods take a more direct approach by integrating all the
knowledge about the system in the posterior pdf. This makes the
results sensitive to the underlying assumptions, but removes strong
systematic biases from the posterior pdf.

\subsection{Extension to 3-D and Monte Carlo integration}
\label{compute}

We can now extend the previous discussion from 1 to 3 dimensions. Let
us consider again the stellar evolution models as a function \F\ 
relating the "physical" parameters $\X\equiv(m,t,z)$ to the
"observational" parameters $\Y\equiv (T, L, \feh)$, with three
components:
\[
\Y={\cal F}(\X)  \Leftrightarrow \left\{ \begin{array}{r c l}
T&=&{\cal F}_T(m,t,z) \\
L&=&{\cal F}_L(m,t,z)\\
\feh&=&\log (z/z_0)\\
\end{array} \right.
\]

Given an observed data triplet,
$$\Y_o\equiv (T_{\obs}, L_{\obs}, [Fe/H]_{\obs})$$
we want to calculate the conditional probability of $\X$, $\Prob(\X|\Y_o)$,
in particular the age conditional probability $\Prob(a|\Y_o)$ for all
possible ages $a$.

According to Bayes' theorem:
\[
\Prob(\X \mid \Y_o) \sim \Prob(\X) \cdot \Prob(\Y_o\mid \X)
\]
Using the marginalization theorem\footnote{the Marginalization theorem
  states that $\Prob(A|B)= \int \Prob(A,C|B) \dd C$}, we integrate over mass
and metallicity, to find the probability that the real age is equal to
a given value $t_0$:
\begin{equation}
\Prob(t_0|\Y_o)\sim \int \!\!\!\int_{R(t=t_0)} \Prob(\X) \Prob(\Y_o|\X) \,\dd m \, \dd z 
\label{margin}
\end{equation}
where the integral is done over the region $R$ defined in the
$(m,t,z)$ space by the condition $t=t_0$.

\Prob($\Y_o \mid \X$) is the likelihood, ${\cal L}(\Y_o,{\cal
  F}(\X))$. For instance, if the uncertainties on the observed
parameters are Gaussian with dispersions $\sigma_{\tiny \log T}$, $\sigma_{\tiny
  \log L}$ and $\sigma_{\tiny [Fe/H]}$, then the likelihood would be as in Eq.~\ref{likelihood}, with $(T,L,\feh)={\cal F}(m,t,z)$.




The term $\Prob(\X)$ is the prior probability distribution. It is the
distribution expected for the parameters a priori, in terms of $m$,
$t$ and $z$.  It can also be thought of as the density in the
$(m,t,z)$ space of an imaginary parent sample, and we shall therefore
note this term like a density, $\rho(\X)$.

In order to compute the integral~(\ref{margin}), the likelihood must
be expressed in terms of $(m,t,z)$. The change of variable from $(T,
L, [Fe/H])$ to $(m,t,z)$ is more complex than in the one-dimensional
case and involves the Jacobian determinant of \F:
   $$ J=
\left|
 \begin{array}{ccc}
  \partial{\cal F}_T/\partial t &\partial{\cal F}_T/\partial m
                                &\partial{\cal F}_T/\partial z\\
  \partial{\cal F}_L/\partial t &\partial{\cal F}_L/\partial m
                                &\partial{\cal F}_L/\partial z\\
  \partial{\cal F}_{[Fe/H]}/\partial t &\partial{\cal F}_{[Fe/H]}/\partial m
                                &\partial{\cal F}_{[Fe/H]}/\partial z\\
 \end{array}
\right|
  $$
  
Then, 
\begin{equation}
\Prob(t_0 | \Y_o) \sim \int \!\!\! \int_{R(t=t_0)} \rho(\X)\, {\cal L} (\Y_o, {\cal F}(\X))\,  J (\X)\, \dd m\, \dd z
\label{margin2}
\end{equation}

In practice, evaluating the Jacobian of the \F\ function at all points
of the 3-dimensional parameter space is a very time-consuming
operation.

Integral~(\ref{margin2}) can be evaluated much more easily by Monte
Carlo integration, which makes the change of variable unnecessary. In
practice, only this approach can ensure results within realistic
computation times for the full 3-dimensional model.
%

%
%
A large sample of $(m,t,z)$ triplets can be drawn following the
density $\rho(m,t,z)$, then the likelihood is computed for all
triplets, and the results collected in age bins, i.e.

\[
\Prob(t_0 \mid \Y_o) \dd t \sim   \sum_{t_0-dt/2 < t < t_0+dt/2}
                             {\cal L}(\Y_o,{\cal F}(m,t,z))
\]

This method has the considerable advantage of requiring no inversion
or differentiation of the \F\ function, which is difficult and subject to
many numerical instabilities, and is even undefined in many regions of
parameter space (where stellar evolution tracks overlap and where no
track passes) and of making the change of variable from $(m,t,z)$ to
$(\feh,T,L)$ easy. It also allows great flexibility as to the
assumptions on the prior. For instance, the inclusion of potential
binarity becomes straightforward (see Section~\ref{binary}).
Another advantage is that building a random sampling of the $\rho(m,t,z)$ 
density is equivalent to the more familiar procedure of generating a synthetic
stellar population, so that existing algorithms designed for this latter task can be
used.

\section{PRACTICAL APPLICATIONS}
\label{Practice}

\subsection{Realistic expressions for the likelihood}

For simplicity, we have made up to now the unrealistic assumption of
Gaussian uncertainties on temperature and luminosity. In practice, the
likelihood can be expressed with suitable assumptions on the
distribution of the uncertainties on the colour, magnitude,
logarithmic temperature or trigonometric parallax. For instance, one
can assume Gaussian uncertainties on $\log T$, $[Fe/H]$ and the
parallax $\parx$.  In that case the likelihood becomes

\[ {\cal L}(T_{obs}, \parx_{obs}, \feh_{obs}, m, t, z) = \]
$$ \frac{1}{\sigma_{[Fe/H]} \sigma_{\parx} \sigma_{\log T} (2\pi)^{3/2}}
 \exp - \frac{(\log T_{obs}-\log {\cal F}_T(m,t,z))^2}{2 \sigma_{\log
     T}^2} $$
$$\times \exp - \frac{({\parx}_{obs}-10^{V_{obs}+2.5 \log{\cal F}_V(m,t,z)/5+1})^2}{2 \sigma_\parx^2} $$
$$\times \exp - \frac{(\feh_{obs}-\log \frac{z}{z_0})^2}{2 \sigma_{[Fe/H]}^2} $$
where $V$ the visual magnitude,
and ${\cal F}_V$ the magnitude predicted from the stellar evolution
models.

\subsection{Choosing a prior}
\label{prior}

Let us now build a realistic prior for the specific case of the solar
neighbourhood.  The prior distributions of $m$, $t$ and $z$ are
assumed to be independent, i.e., no prior information is assumed on an
age-metallicity relation, or on a time variation of the mass
distribution.  In that case, the mass, age and metallicity prior can
be considered separately:
\[
\Prob(\Y)=\Prob(m)\cdot \Prob(t) \cdot \Prob(z)
\]

\subsubsection{The mass prior}

The mass prior can be chosen as one's favourite initial mass functions
(IMF) derived for the Galaxy. Within reasonable limits, the precise
choice of IMF will not have a strong influence, because the mass range
covered by the F and G dwarfs is not large.

\subsubsection{The age prior}

The age prior is the expected age distribution of all stars ever born
in the sample considered (the fact that some of them have already died
is accounted for in the \F\ function), in other words the star formation
rate (SFR) of the sample considered. The SFR of the Galactic disc is
not precisely known. It seems to have been globally constant or
slightly decreasing (Hernandez, Valls-Gabaud \& Gilmore 2000; Chang,
Shu \& Hou 2002; Vergely et al. 2002), but its small-scale structure
is still largely unknown. At this stage a flat age prior is a
reasonable assumption. Decreasing priors can also be used. Within
reasonable limits, the slope of the SFR does not make large
differences in the recovered age pdfs.

\begin{figure}
\caption{
  Magnitude probability distribution corresponding to a flat age
  prior, with $\log T=3.76$ and $\feh=0.0$. The sharp shape is due to
  the highly non-linear nature of the function ${\cal F}_L(m,t,z)$.  }
\resizebox{\hsize}{!}{\includegraphics{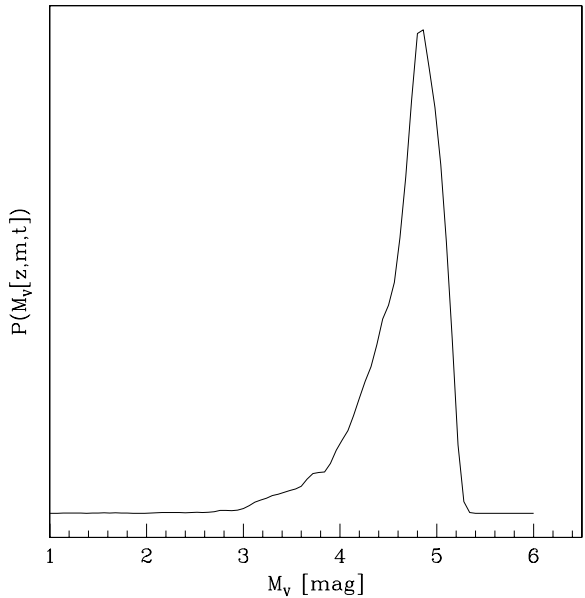}}
\label{ageprior}
\end{figure}

Note that  using a flat age prior is not at all equivalent to ignoring 
the age prior. A flat prior in age is far from translating into a flat prior
in parameter space (see Section~\ref{bayesth}). 
 Fig.~\ref{ageprior} shows the prior distribution of magnitude at solar values
 of temperature and metallicity resulting from a flat age prior. The abrupt slope
towards bright magnitudes is due to the acceleration of evolution in
temperature and magnitude after the main-sequence phase. The cutoff at
faint magnitudes is obviously due to the absence of models below the
zero-age main-sequence.

For a flat or slightly decreasing age prior, an upper age cutoff must
be chosen. This somewhat arbitrary procedure has a direct influence in
the derived age distribution by simply removing all ages above the
cutoff. At present, the maximum age of the stars in the Galactic disc
is not well known. The age of the thin disc has been studied by Binney
et al. (2000), but the solar neighbourhood also contains thick disc
stars, whose age may be several Gyr higher than that of the oldest
thin disc stars.

\subsubsection{The metallicity prior}

\begin{figure}
\caption{Metallicity distribution for the solar neighbourhood and
  metallicity prior. {\bf Stars:} G-dwarf volume-limited metallicity
  distribution according to J{\o}rgensen (2000). {\bf Dotted line:}
  Metallicity distribution of the \GC\ survey. {\bf Dashed line:}
  Metallicity prior adopted in the present paper. {Insert:} Gaussian
  probability distributions corresponding to dispersions
  $\sigma_{[Fe/H]}=$0.05, 0.10 and 0.15 dex.}
\resizebox{\hsize}{!}{\includegraphics{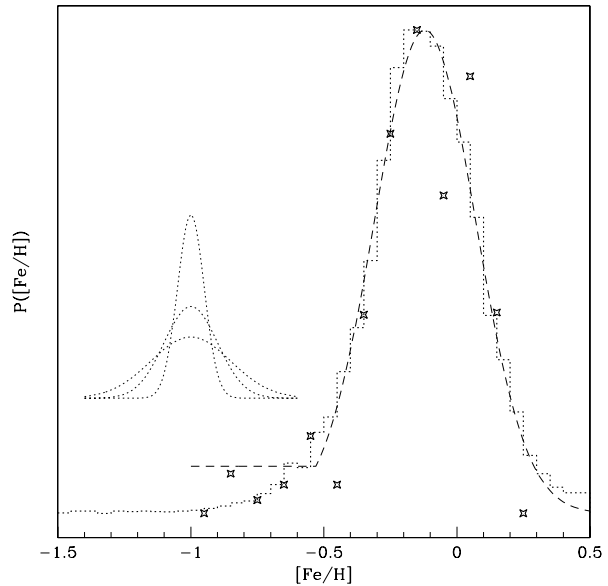}}
\label{metalprior}
\end{figure}

In the case of the Galactic disc, a good metallicity prior is the
expected distribution of a volume-limited sample of the solar
neighbourhood. Two such distributions are displayed on
Fig.~\ref{metalprior}. The influence of the metallicity prior depends
on the size of the observational uncertainties on $\feh$. For very
high accuracy spectroscopic metallicities, with $\sigma_{[Fe/H]}\simeq
0.05$ dex (e.g. E93), the likelihood is narrow enough to overwhelm the
changes in the prior. For uncertainties most typical of larger surveys
and of photometric metallicities, $\sigma_{[Fe/H]}\simeq 0.10$ dex,
the influence of the prior becomes more significant, especially in the
regions where it is varying more rapidly: at the high-metallicity end
and at the connection between the main thin-disc distribution and the
thick-disc tail near $\feh \simeq -0.5$.
The systematic bias on likelihood-based methods can reach 0.1 dex in
metallicity, causing systematic biases on the derived ages.

When the metallicity uncertainty $\sigma_{[Fe/H]}$ is even higher, for
instance when collecting metallicities from different calibrations
(e.g. Ibukiyama \& Arimoto 2002, with $\sigma_{[Fe/H]}\sim 0.15$ dex),
the influence of the metallicity prior will become so important that
the derived ages will be highly dependent on it and extremely
uncertain. Maximum likelihood ages will be strongly biased, and will
produce visibly unreliable results such as fig.~5 of Ibukiyama \&
Arimoto (2002). It is apparent from our Fig.~\ref{metalprior} that
$\sigma_{[Fe/H]}\sim 0.15$ implies that the variation of the
likelihood will not be steeper than the variation of the prior, which
is the Bayesian definition of a "bad measurement".

\begin{figure}
\caption{Theoretical isochrones from the Padua models for 
  $\feh=-0.7$ (0,1,2,3,5,10 and 15 Gyr) and objects in the \GC\ 
  catalogue with $-0.75<\feh<-0.65$. Detected binaries are indicated
  with open symbols. The error bars show the uncertainties due to the
  parallax. The dashed line is the solar-metallicity ZAMS. Part of the
  obvious temperature mismatch between models and observations may be
  due to the slope of the metallicity prior. Taking the models and
  observations at face value would lead to assigning terminal ages to
  practically all stars.}
\resizebox{\hsize}{!}{\includegraphics{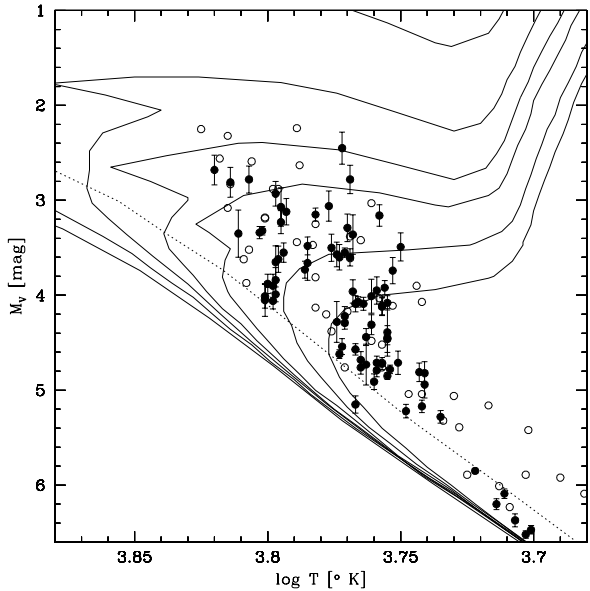}}
\label{metal}
\end{figure}

\subsection{Accounting for undetected binaries}
\label{binary}

Up to this point, we have considered that all stars in our sample
obeyed the \F\ relation between $(m,t,z)$ and $(T, L, \feh)$ with
Gaussian uncertainty distributions. Even without considering such
second-order effects as rotation or Helium abundance, this relation
does not always hold for real stars, particularly in the case of
undetected binaries.  The light from the companion of an undetected
binary can move a given object up to 0.75 mag (for equal-mass
binaries) above its true position in the colour-magnitude diagram.
This obviously has a profound effect on the age determination.

If the number of undetected binaries in the sample is not too large,
its effect on the age determination for main-sequence stars will be
manageable. A few of the ages will turn out to be overestimated.

The effect of the binaries on the age determination of {\it evolved}
stars in the subgiant zone, however, will be large. Because the
evolution is more rapid in the subgiant zone, the probability of
finding a star in a subgiant stage is much lower than for the
main-sequence stage. On the other hand, undetected binarity can move
main-sequence objects upwards into the subgiant zone of the $(T,L)$
plane. As a result, the contamination of binaries in the subgiant zone
can be very important.

The standard approach to the age determination offers no obvious way
to deal with the binary contamination, which has to be treated as a
nuisance and studied with separate simulations.

In the Bayesian formulation, the inclusion of undetected binaries is
not particularly difficult. A term can be added to
equation~(\ref{margin}) integrating the hypothesis that the object can
be an undetected binary:
\[
\Prob(t|\Y_o)\sim \Prob(t) \cdot \Prob(\Y_o|\text{\textrm{single or binary}, t}) 
\]
Because binarity and non-binarity are mutually exclusive, the
probability sum rule can be used to yield:
\[
\Prob(t|\Y_o)\sim \Prob(t, \textrm{single}) \Prob(\Y_o|t,\textrm{single})
  \]
$$+ \Prob(t, \textrm{binary}) \Prob(\Y_o|t, \textrm{binary})$$
If age and binarity are independent, then
\[
\Prob(t|\Y_o)\sim \Prob(t) ( \Prob(\Y_o|t, \textrm{single}) + q\, \Prob(\Y_o|t, \textrm{binary}))
\]
where $q\equiv \Prob(\textrm{binary})/\Prob(\textrm{single})$ is the
rate of undetected binaries.

In practice, because binarity has such a large effect on the age
determination, one may be less interested in knowing the age pdf in
the case of binarity than to know the total probability for the star
to be a binary, "$\Prob(\textrm{binary}\mid \Y_0)$". To calculate this
probability, we integrate over all values of~$t$:
\[
P_{bin} \equiv \Prob(\textrm{binary}|\Y_o)=\]\[
 \int \Prob(\textrm{binary}, t|\Y_o)\dd t =\]
$$ q  \int \Prob(t) \Prob(\Y_o|t, \textrm{binary}) \dd t $$

The terms inside the integral can be calculated as in
Section~\ref{theory}, using the modified relation ${\cal F'}$ between
$(m,t,z)$ and ($T, L, \feh$) suitable for binaries, depending on the
mass ratio parameter.

The computations show that $P_{\mbox{\scriptsize bin}}$ is small in
some parts of the $(T,L)$ plane and much larger in others,
particularly 0.75 mag above the main-sequence, as could be expected.
At that position, it reaches about 10 times the value of $q$ (implying
that for an undetected binary rate of 10 percent, the object is
actually as likely to be a binary as a single star). The interesting
thing is that the Bayesian computation not only yields a specific
value of $P_{\mbox{\scriptsize bin}}$ for each object for any
subsequent statistical study, but also includes in the posterior age
pdf the possibility of the star being an undetected binary. In this
way, undetected binaries are less likely to introduce unrecognized
contamination in the scientific analysis of the results (see
Section~\ref{model}).

\subsection{Choice of stellar evolution models and temperature scale match}
\label{tempshift}

Sets of stellar evolution tracks for late-type dwarfs have been
produced by many different groups.  Some of the most widely used are
Girardi et al. (2000), Yi et al. (2001), Lejeune \& Schaerer (2001).
The agreement between the predicted isochrones from the different
groups is generally good on or near the main sequence, so that using
one set of models rather than another doesn't introduce dramatic
differences in the derived stellar ages.  Two robust predictions of
stellar evolution theory, that heavier stars evolve more quickly and
that stars on the main sequence get brighter with age, provide the
dominant tendencies.

There are, however, important residual differences between the sets of
models, that have a significant influence on the age determination. Of
particular importance are the known difficulties related to the model
temperature predictions: absolute temperature scale of the models,
colour-temperature conversions, metallicity dependence of the position
of the unevolved main sequence. There also are significant systematic
differences between the model temperature predictions and the observed
position of well-measured field dwarfs (Lebreton et al. 1999; Lebreton
2000; Kotoneva et al. 2002b). 

As far as the age determination is concerned, the important fact is
that the models and observations be on the same {\it relative}
temperature scale. The observation of nearby unevolved K-dwarf stars
with well-known parallaxes and metallicities shows that the actual
colour change with metallicity is significantly lower than model
predictions (Kotoneva et al. 2002b). Several explanations have been
proposed for this mismatch, including problems with the
temperature-colour conversions, metallicity-dependent helium
abundances, and heavy-element sedimentation (Lebreton et al. 1999).
None of these effects seems able to account for the whole mismatch.
Some moderately metal-deficient dwarfs in the solar neighbourhood are
compared with model predictions in Fig.~\ref{metal} to illustrate the
amplitude of the mismatch between the Girardi et al. (2000) models and
the observations. Obviously, ages derived with such a large
temperature mismatch will be strongly biased towards terminal ages.

Several schemes can be adopted to ensure that the models and
observations are coherent, either at the level of the
colour-temperature conversion, or as empirical temperature shifts in
the models.

%
%
%
%
%
%
%
%
%
%
%

\section*{PART TWO: Application to the E93 sample and the
                     age-metallicity relation}

\section{Bayesian ages for the E93 sample and
                       solar neighbourhood AMR}

\label{e93}

In this Section, we apply the Bayesian age calculation to a specific
case, the landmark E93 study (see Introduction).  As an illustration
and important application of the approach developed in the previous
sections, we calculate the posterior age pdf for the objects of the
E93 sample, and reconsider their determination of the age-metallicity
relation of the Galactic disc.


\subsection{Recent data for the E93 sample}

The E93 sample was selected from the large Olsen (1994) catalogue of F
and G dwarfs in the solar neighbourhood. The selection criteria were
approximately a range in temperature, $5600 < T < 6800$, and in
evolution away from the Zero-Age Main Sequence, $M_V-M_{V,ZAMS} > 0.4$ mag. The
objective was to select stars in the subgiant portion of the CMD,
where the isochrones are more widely spaced and age determinations are
presumably more accurate.

The main emphasis of E93 was on providing accurate metallicities. They
estimated the relative accuracy of their metallicities to be 0.05 dex.
They derived ages for their objects by comparison with Vandenberg
(1985) isochrones. The adopted ages were that of the isochrone
crossing the position of the data in the temperature-luminosity plane.
The uncertainties on the ages were estimated to be around 0.1 dex,
based on the direct propagation of the observational errors.  In our
notation, E93 have computed age estimates from
\[
t={\cal F}^{-1} (T, L, \feh)
\]
and evaluated the uncertainties with
\[
{\cal F}^{-1} (T\pm \sigma_T, L\pm \sigma_L, \feh \pm \sigma_{[Fe/H]})
\]

E93 provide evidence for the 0.1 dex value of the age uncertainties
by displaying data for M67 and showing that, indeed, the dispersion of
the inferred ages is of the expected order, at least in the subgiant
portion of the CMD.

The data for all E93 objects have been significantly improved in the
intervening decade, the only exception being the individual
metallicities, that were of high relative accuracy in E93. The most
noteworthy addition is the availability of Hipparcos parallaxes, that
allow the distances -- hence the absolute magnitudes -- of the objects
to be known with much better accuracy than was available to E93. Most
stars in the sample have distances of less than 50~pc, and
correspondingly uncertainties of less than 5 percent in the Hipparcos
parallaxes, in contrast with the 14 percent uncertainty assumed by E93
for the photometric distances.

The second important addition to the E93 data is that of further
radial velocity monitoring (\GC), that has revealed a certain number
of spectroscopic binaries. 

Ng \& Bertelli (1998) have reconsidered the E93 sample with Hipparcos
distances, and with the Bertelli et al. (1994) stellar evolution
models. However, their age derivation method is not fundamentally
different from E93, and they consequently derive a similar
age-metallicity plot. It is already apparent in their study, though,
that the new distances considerably reduce the number of stars in the
high-metallicity, high-age section of the diagram. This could be
expected, because it is precisely in this region, the terminal-age
subgiant region, that objects get preferentially scattered by the
distance errors (see Section~\ref{laurent}). It is also worth noting
that many stars are put back on the main-sequence with $\Delta M_V <
0.4$ by the new distances, proving the reality of the biases
associated with the non-Bayesian calculation.

Another valuable improvement was brought by Lachaume et al. (1999),
who computed the distribution of the likelihood for the age estimation
of a sample of 91 local field dwarf stars. While this still ignores
the prior pdf, it does show that the 1-$\sigma$ interval for the
derived ages is larger than 0.1 dex for many objects in
the crucial 3-10 Gyr range.

Finally, the Hipparcos data have also allowed the discovery of large
temperature shifts between models and data (see
Section~\ref{tempshift}), that were not corrected by E93 or Ng \&
Bertelli (1998) and also affect the age determinations.

\subsection{Bayesian ages for the E93 sample}

\label{agese93}

Posterior age pdf's were computed for the objects in the E93 sample
using the method of Section~\ref{compute}. The age prior is taken as
flat, with a cutoff at $t_{max}=15$ Gyr, the mass prior as a power
function of slope $-2.35$. The metallicity prior is a Gaussian
centered on $\feh =-0.15$ with a dispersion of 0.19 above
$\feh=-0.53$, and a constant below $\feh=-0.53$ with a cutoff at $\feh=-1.0$ (see
Fig.~\ref{metalprior}). This function is a visual approximation of the
metallicity distribution of the whole \GC, and is also compatible with
the volume-limited distribution for the solar neighbourhood derived by
J{\o}rgensen (2000). The stellar population synthesis code IAC-star
(Aparicio \& Gallart 2004) was used for the Monte-Carlo estimation of
$\rho(m,t,z)$ and the ${\cal F}$ transformation.  The E93
metallicities were put on the scale of Santos et al. (2002) by a shift
of +0.12 dex. The temperature scale in \GC\ (as of 2003) was used, adjusted by a
shift of $+0.006$ to obtain a satisfactory match between the Padua
isochrones and the \GC\ data in the theoretical plane\footnote{ The discussion of
  this mismatch is beyond the scope of this article, but it is
  certainly worth enquiring into and is a strong limitation on the accuracy
  of the isochrone age estimates (see review by Lebreton 2000 and
  Section~\ref{tempshift}).}.

Gaussian uncertainties were assumed on $\feh$, $\log
T$ and $M_V$. 
Several different sets of values were used for the standard
dispersions of the uncertainties. See Section~\ref{model} for a
confrontation of different cases. For our "standard" computation, we
used $\sigma_{\log T}=0.009$ $\sigma_{Mv}=0.15$ and
$\sigma_{[Fe/H]}=0.075$.  We arrive at these values by using the
uncertainties proposed by Ng \& Bertelli for their revision of the E93
sample ($\sigma_{[Fe/H]}=0.05$, $\sigma_{\log T}=0.006$), an $M_V$
uncertainty of 0.10 mag (5 percent uncertainty on the distance), and
allowing for the possible presence of systematic errors of similar
amplitude by multiplying all values by a factor 1.5. It is important
to remember that differences such as a zero-point shift in the
metallicity scale or temperature scale have a strongly non-linear
effect on the age determination (i.e. shifting the temperature scale
does not produce a single shift of all age values but very different
shifts depending on the position in the CMD). Systematic zero-point
differences of 0.10 dex in $\feh$ and 0.005 in $\log T$ are common
between different scales. Indeed, as mentioned above, shifts of such
magnitude were found necessary to match the theoretical isochrones
with the observational data. The use of $\feh$ itself as a surrogate for the
total heavy-element abundance used in the theoretical models is also
subject to uncertainties, given the observed variations in the
abundance ratios from star to star.  In the Bayesian treatment, these
sources of error have to be integrated into the assumed observational
uncertainties. It is crucial to use the real difference between the
data and the evolution tracks to estimate the likelihood, not the
relative difference.

\begin{figure}
\caption{Posterior age pdf, between 0 and 15 Gyr, for representative
  late-type examples of the updated E93 data. The dotted line indicates
  a Gaussian of 0.1 dex dispersion around the E93 age value, for
  comparison.}
\resizebox{\hsize}{!}{\includegraphics{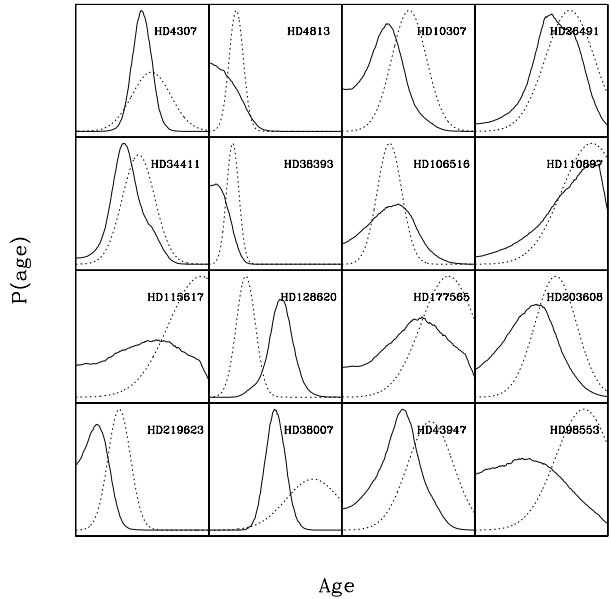}}
\label{voir}
\end{figure}

Fig.~\ref{voir} gives posterior age pdf for a few representative
objects in the sample, compared with a 25 percent ($\sim 0.1$ dex) dispersion Gaussian
centered on the E93 age estimate.  The results show that the shape and
width of the posterior age pdf can vary a lot from one star to the
next. Both the central value and the general shape of the posterior
probability distributions of the age are often very different from
that obtained by E93 with the ${\cal F}^{-1}$ approach. In some cases,
a Gaussian is a valid approximation, but many stars
are subject to wider and very asymmetrical probability
distributions. Some have posterior pdf spanning most of the allowed
0-15 Gyr range, so that the chosen age value is sensitively dependent
on the assumed prior. In these cases, the derived age is not well
determined (e.g. HD~115617, HD~177565, HD~98553 in the Figure).

\subsection{Age-metallicity relation}

\begin{figure*}
\resizebox{8cm}{!}{\includegraphics{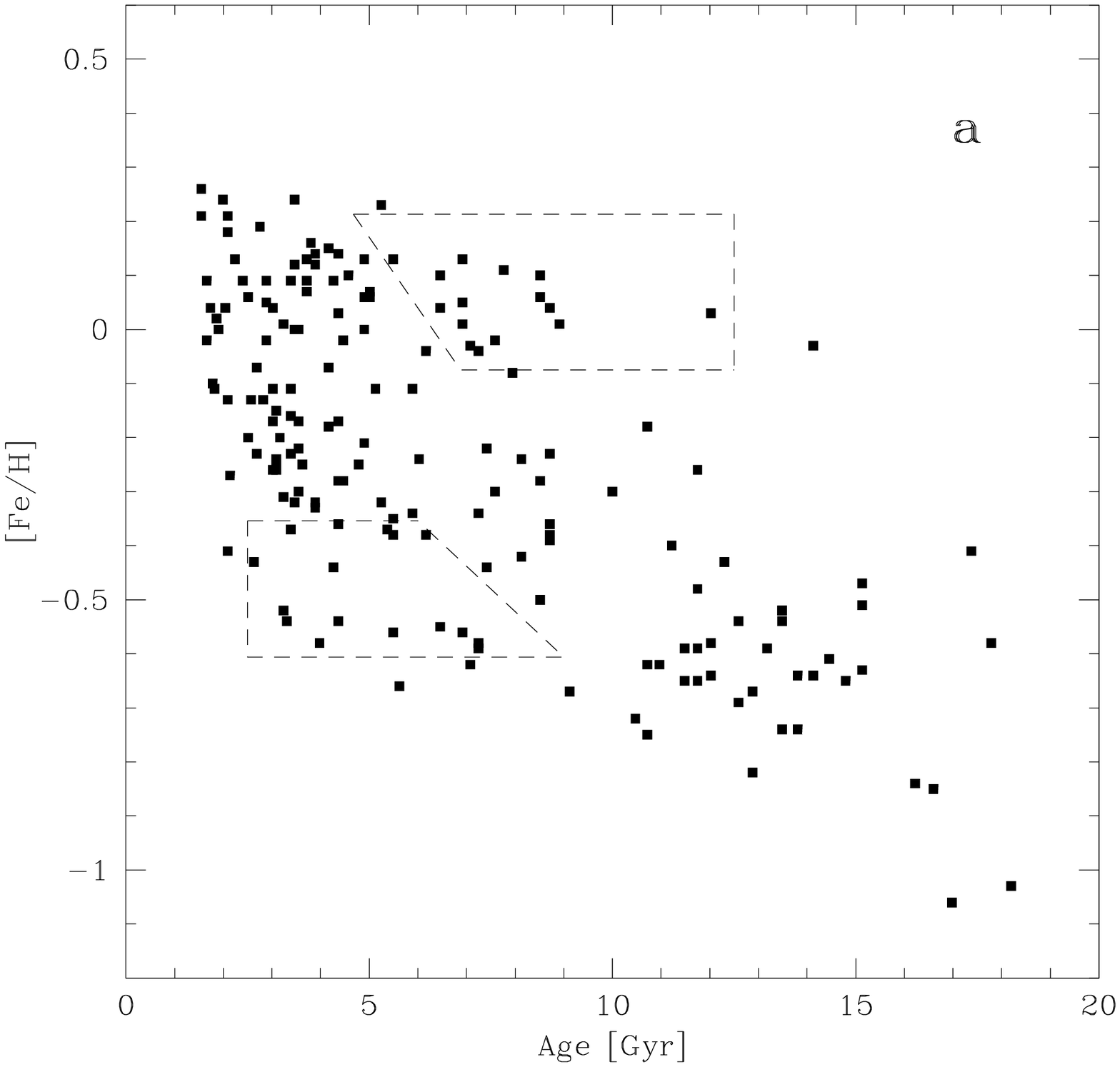}}
\resizebox{8cm}{!}{\includegraphics{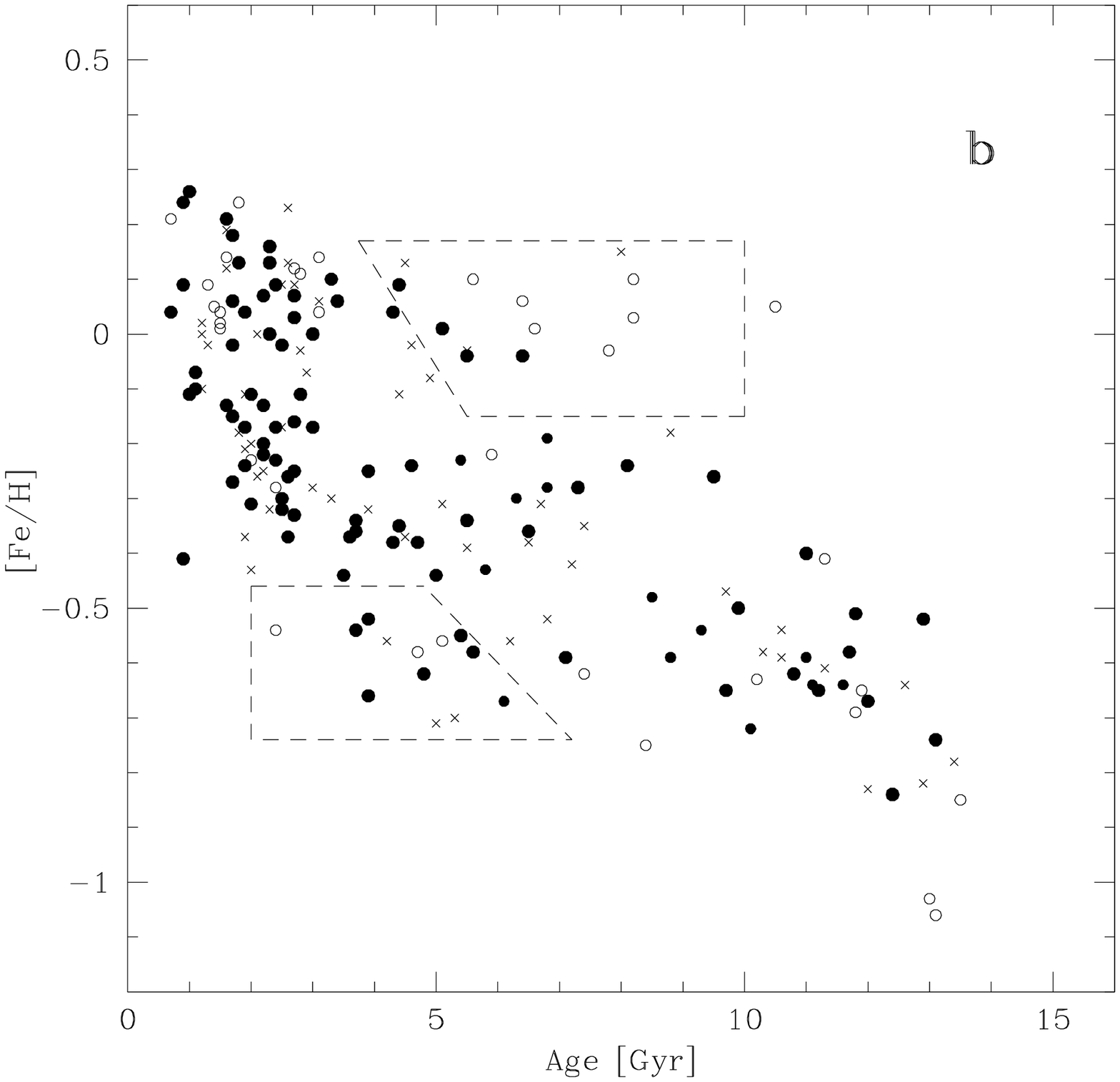}}\\
\resizebox{8cm}{!}{\includegraphics{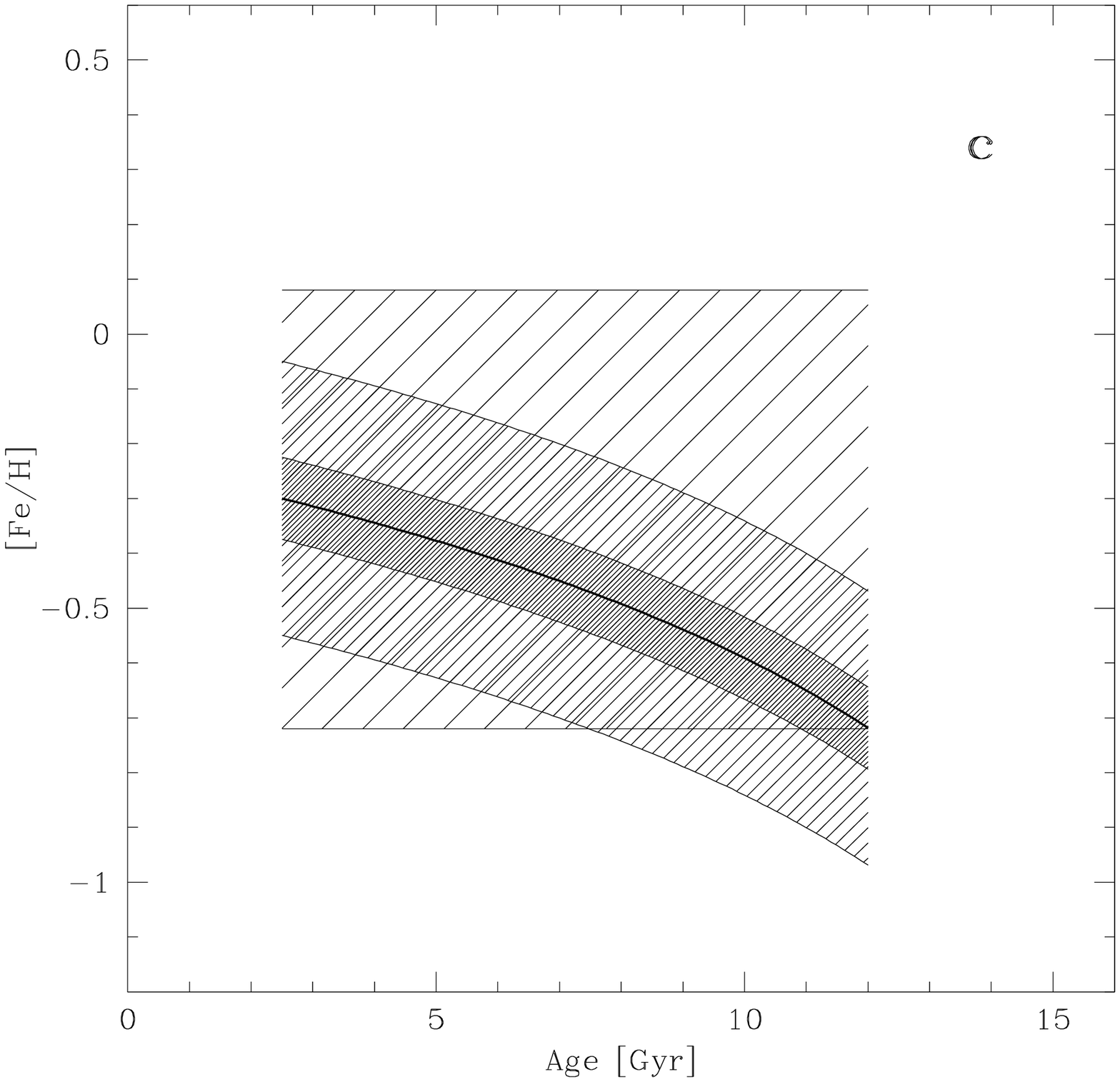}}
\resizebox{8cm}{!}{\includegraphics{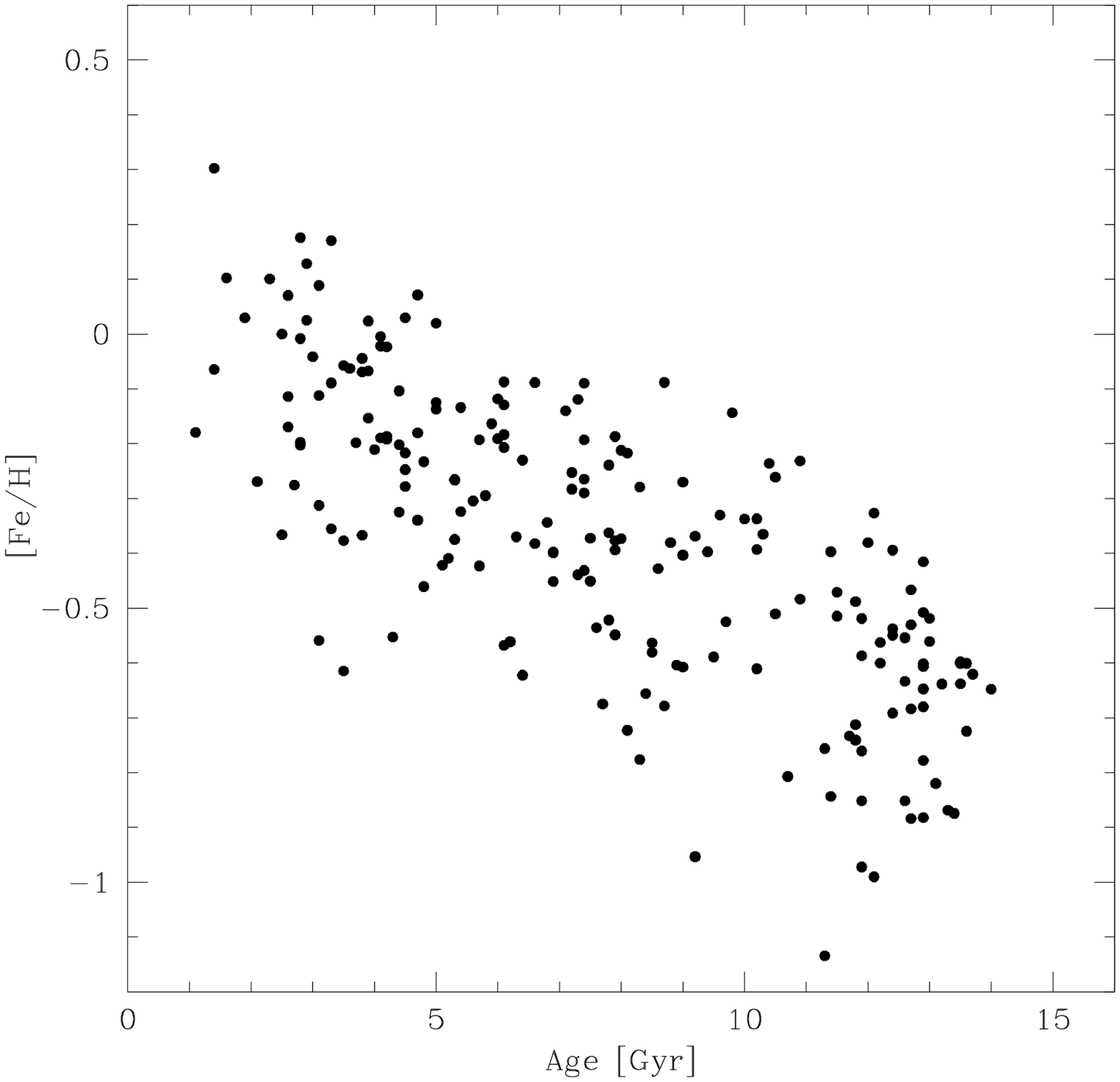}}
\caption{{\bf a}
  The original age-metallicity plot from E93. {\bf b} the
  age-metallicity plot for the E93 sample with new data: luminosities
  from Hipparcos distances, temperatures from \GC,
  ages from the present study. Symbols as follows: {\it
    crosses}--binary in the \GC\ catalogue, {\it open
    circles}--photometric distance from E93 discrepant with Hipparcos
  distances by more than 20 percent, metallicity from E93 and from
  \GC\ by more than 0.2 dex, or temperature/colour discrepant between
  E93 and \GC.  {\bf c} The age-metallicity relations used in the
  Bayesian models of Section~\ref{model}. {\bf d} Simulated
  age-metallicity plot, resulting from an age-metallicity relation of
  total range $\Delta [Fe/H]=0.4$ dex, and observational uncertainties
  $\sigma_{[Fe/H]}=0.05$, $\sigma_{T}= 75$ K, $\sigma_{M_V} = 0.15$
  mag, systematic shifts 0.12 dex in metallicity and +0.005 in $\log
  T$. In the two top plots, the dashed lines outline two regions that
  are important for the evaluation of the dispersion of the
  age-metallicity relation, and where contamination from biased
  low-accuracy ages is expected to be significant (see text).
  "Region~I" near solar metallicity, age$>5$ Gyr, "Region~II"
  metal-poor, age$<6$ Gyr. }
\label{amr_e93}
\end{figure*}

Fig.~\ref{amr_e93}b plots the Bayesian ages, with the updated data, in
the age-metallicity plane, using the median of the posterior pdf as an
age estimator. The binaries identified in \GC\ are indicated as
crosses.

Given the sensitivity of the age determination to the input
parameters, it is important to use all the available information to
identify possible outliers. The metallicities of E93 were confronted
with the photometric metallicities derived in \GC, the Hipparcos
parallax distances with the distances derived from the photometric
calibrations in \GC. The following conditions were required for
inclusion in the final sample:
\begin{eqnarray*}
\mid \feh_{E93}-\feh_{\GC} \mid < 0.2 \\
\mid \frac{r_{E93}-r_{\GC}}{r_{E93}} \mid < 0.2 \\
\end{eqnarray*}
where $r$ is the distance. HD 67228, 112164, 199960 and 207978 were
identified visually as outliers on a comparison of temperatures
between E93 and \GC, and of colour between the Str\"omgren $(b-y)$ and
Hipparcos $(B-V)$.

The objects singled out by the conditions above are indicated by open
symbols in Fig.~\ref{amr_e93}b. The data discrepancy indicates that
they can be peculiar in some way, or that some of their measurements
may be statistical outliers, so that the age determination may be
affected.

Fig.~\ref{amr_e93}a repeats the original E93 age-metallicity plot, for
comparison. Two regions particularly important in giving an impression
of high dispersion in the age-metallicity relation, and particularly
prone to contamination by skewed probability distributions of age with
large error bars (see Section~\ref{compute}), are indicated with
dashed lines (Similar regions in the age-metallicity plot were
previously used by Rocha-Pinto et al. 2000 to show that ages from
chromospheric activity did not produce a high intrinsic dispersion in
the AMR).  The presence of many stars in these two regions in E93
contributed strongly to the conclusion of a very wide metallicity
dispersion at all intermediate ages.

\subsection{Discussion}

\begin{figure}
\caption{Median Bayesian age in the age-metallicity plot for our
  selection of the E93 sample. For a few objects, the
  full-width-half-maximum and the full width at tenth maximum of the
  posterior age pdf are indicated. The solid and dashed lines indicate
  respectively the mean and envelope AMR used for Fig.~\ref{mmr}.}
\resizebox{\hsize}{!}{\includegraphics{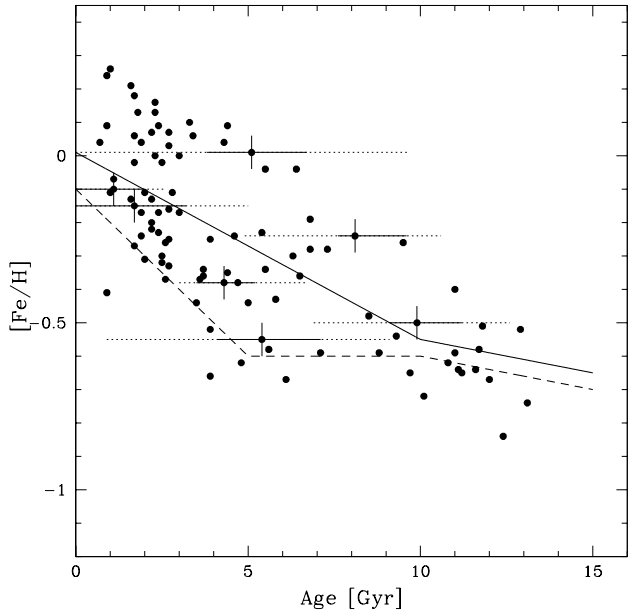}}
\label{amr}
\end{figure}

Fig.~\ref{amr_e93} offers a spectacular confirmation of the effect of
biases and of the potential perils of replacing the complete posterior
pdf by a single maximum-likelihood point in the age-metallicity plot.
Indeed, the new data give a strong indication that there were many
objects in the high-age, high-metallicity part of the original E93 AMR
plot that had been displaced in the subgiant zone by high
observational uncertainties and binarity ("terminal age bias", in our
terminology). The upper dashed zone in the AMR becomes practically
empty with the new data. All four points remaining in it are
compatible with being one-sigma outliers from younger ages.

Fig.~\ref{amr} displays the age-metallicity diagram for our selected
objects, plotting the half-maximum and tenth-maximum intervals of the
posterior age pdf for a few representative objects to give a feeling
for the shape of the age pdf.  Contrarily to the original E93 AMR, the
updated AMR diagram outlines a definite monotonic relationship between
age and metallicity for intermediate ages. Part of the scatter in the
original relation is removed. A simple linear fit gives $\feh=-0.056\,
\mage +0.011$, with a dispersion of 0.18 dex in metallicity, to be
compared with 0.24 dex in E93. This value still includes the scatter
introduced by the age uncertainties and by the fact that the actual
AMR may not be linear\footnote{as well as the increased scatter
  introduced by the deliberate selection by E93 of objects of
  different metallicities and masses. See their discussion on this
  point.}. Therefore, 0.18 dex can be considered a strict upper limit
for the "cosmic scatter" in the AMR.

Using the posterior age pdf that we obtain for the E93 objects, Monte
Carlo simulations of the whole procedure were carried out to determine
what dispersion is expected from the observational uncertainties
alone, assuming a dispersionless AMR. We find that a dispersion at
fixed age $\sigma_{[Fe/H]} \simeq$0.10 is introduced around a
dispersionless AMR by the observational uncertainties. By
quadratically subtracting this dispersion from the value of 0.18 found
in the observed AMR, we estimate the remaining intrinsic scatter, the
so-called "cosmic scatter", to be at most 0.15 dex.

Therefore, with the improved data and detailed treatment of the age
probability distribution, the E93 sample no longer indicates a very
high scatter in the metallicity at a given age, or a near-absence of
AMR in the intermediate age range.  While still clearly distinct from
a dispersionless AMR, the data indicate a rather well-defined growth
of mean metallicity with time, with an intrinsic dispersion of the
order of 0.15 at most.
 
A similar conclusion was already obtained by Rocha-Pinto et al. (2000)
on the basis of chromospheric ages for solar-neighbourhood dwarfs.
Their result was subsequently criticized on the basis of the observed
disagreement between isochrone ages and chromospheric ages (e.g.
Feltzing et al. 2001). However, the present study brings support to
the validity of the indications from chromospheric ages and shows how
the apparent disagreement could arise from strong systematic errors on
the isochrone ages.
 
In the following two paragraphs, we examine other ways to look at the
data that can add more indications on the reality of the high
intrinsic scatter, by considering the two regions, "Region I" and
"Region II", that we defined in the age-metallicity diagram on
Fig.~\ref{amr_e93}.

\begin{figure}
\caption{Mass-metallicity relation for our sample. Symbols as in
  Fig.~\ref{amr_e93}b. The thick lines indicate the lower envelope of
  the mass-metallicity relation for a low-dispersion AMR (solid line
  in Fig.~\ref{amr} with a 0.24 dex range in [Fe/H]) and high
  dispersion AMR (dashed line).  The oblique dotted and solid lines
  indicate the zones affected by the temperature selection biases.}
\resizebox{\hsize}{!}{\includegraphics{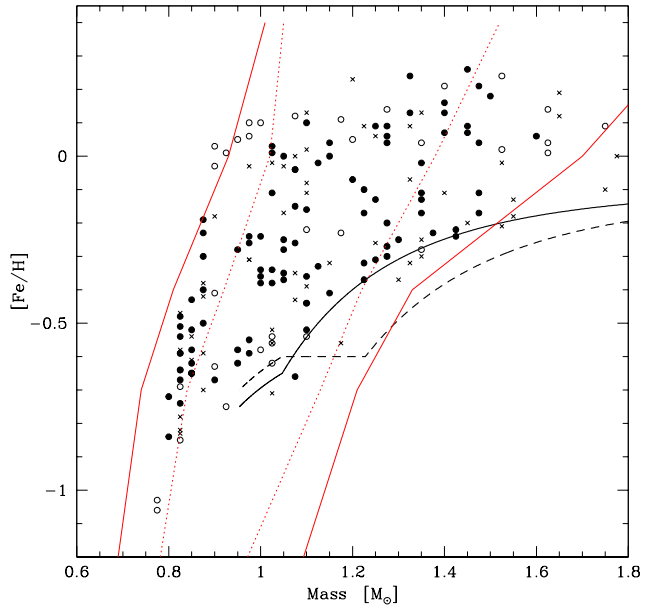}}
\label{mmr}
\end{figure}

\subsection{"Region I": solar metallicity, age $>5$ Gyr}
\label{regions}

The majority of the objects placed in or near "Region I" of the CMD
have been excluded by our quality criteria. There are five detected
binaries. The other objects are stars with $M\sim 0.9 M_\odot$, for
which the Hipparcos distance is much smaller than the distance
obtained by E93 from the photometry (more than 20 percent difference,
these objects are HD~76151, 86728, 108309, 115617, 127334, 177565 and
217014). Because the main-sequence is narrow for $M\sim 0.9 M_\odot$,
these smaller distances, implying fainter $M_V$, are sufficient to
bring these stars from the subgiant zone (with apparently well
determined ages in the 5-10 Gyr range) back on the main sequence.
These objects were therefore outliers of the Str\"omgren $M_V$
calibration that had been preferentially selected by the $\Delta
M_V>0.4$ criteria of E93. The Hipparcos distances put them back on the
main-sequence where, for such low masses, no accurate estimate of the
age can be given.

This illustrates a side-effect of the $\Delta M_V>0.4$ selection
criteria used by E93. Intended to select only stars in the region of
the CMD were isochrones are well-spaced, it also samples the region of
the CMD where the difference between likelihood and posterior is
larger, and the proportion of undetected binaries is higher. The E93
study picked up 189 stars out of more than 19'000, and in the process
it favors the 2-3 $\sigma$ outliers from the main sequence. Simple
statistical considerations show that the number of main sequence
contamination in the $\Delta M_V>0.4$ zone must be large. The Bayesian
approach automatically includes this selection effect in the posterior
pdf, by taking into account the prior distribution $\Prob(t)$, which
is heavily weighted towards the main-sequence,
and the bias caused by the selection criteria is absent in our updated
sample, thanks to the fact that the Hipparcos parallaxes are used to
re-compute values of $M_V$ that are independent of the original
photometric $M_V$ used in the sample selection.

Note that "Region I" is even more populated in the higher-uncertainty
samples of Ibukiyama \& Arimoto (2000) and Feltzing et al. (2001). The
presence of many objects in this region of the age-metallicity diagram
does not give any indication of the real existence of such
high-metallicity, intermediate-age stars.  On the contrary, the E93
sample tends to indicate that this region is actually empty, because
there is not a single bone fide subgiant with an age above 7 Gyr and a
metallicity near solar.

Interestingly, the Sun, at $t=4.5$ Gyr and $\feh$=0, seems to be near
the upper edge of the compatible age-metallicity distribution of the
sample (Wielen et al. 1996). This may be related to the observed
statistical overabundance of planet host stars (Gonzalez 1998; Santos,
Israelian \& Mayor 2000).

\subsection{"Region II": metal-poor, age $<6$ Gyr}

Let us now consider the second dashed region, "Region II", in the
low-age ($t\leq 6$ Gyr), low-metallicity ($\feh \leq$ 0.5) part of the
diagram.

The age-metallicity diagram shows a few data points in this region
that may or may not have been scattered there by the uncertainties on
the age determination (for none of these objects is the whole
posterior age pdf entirely contained in "Region II"). Fortunately,
there is another way to determine whether this region of the AMR is
really occupied.

The age estimation from isochrones also provide an estimation of mass.
The mass can be derived from theoretical tracks in a more reliable way
that the age, because the mass changes relatively slowly with the
observed parameters. A fundamental feature of stellar evolution is the
fact that the duration of the main-sequence phase is a very sharp
function of mass. This sharp dependence implies the following: only
stars below a certain mass can reach ages above a given age. E.g., only
stars with masses below 1.2 $M_\odot$ can reach ages above 5 Gyr, and
masses below 1 $M_\odot$ ages above 10 Gyr. This relation between mass
and maximum age implies that the {\it lower envelope} of the
mass-metallicity relation will depend on the AMR and its dispersion.
As we move to lower masses, higher ages become available and the
metallicities reached at these ages begin to appear in the age-mass
relation.

Therefore, if there really are low-age (3-6 Gyr), low-metallicity
($\feh\leq -0.5$) stars --objects within Region~II-- then we expect
such stars to be of all masses able to reach at least 3 Gyr of age,
$M< \sim 1.3 M_\odot$. On the other hand, if Region~II is not actually
occupied, and the points in the age-metallicity diagram are scattered
into it by the uncertainties from higher ages, then this should be
revealed by the absence of $M \sim 1.1 - 1.3 M_\odot$ stars with lower
metallicity. The region in the mass-metallicity diagram from which
low-metallicity stars start to be found gives a definite indication of
the minimal age of such stars.

Fig.~\ref{mmr} shows the mass-metallicity plot for the E93 sample with
the updated data. E93 do not give mass estimates for their stars.
Masses for Fig.~\ref{mmr} were computed by us as a by-product of the age computation.
 The upper envelope of
the relation between mass and maximum age was adjusted on the \GC\ 
data as $\log t_{lim}= 1.09-4.35 \log (mass)$. Via this relation, the
lower envelope of the AMR can be converted into a lower envelope in
the mass-metallicity plot.  Fig.~\ref{mmr} gives the predicted lower
envelopes of the mass-metallicity relation with the two AMR plotted in
Fig.~\ref{amr}. The first is a $\Delta=0.24$ dex relation
defined by the solid line on Fig.~\ref{amr}, and the second (dashed
line) is the lower envelope of an AMR with very high intrinsic
scatter, of the type inferred from the original E93 interpretation.
The crucial difference between the two AMR is that one predicts the
real existence of objects in Region II while the other does not. The
zones affected by the selection biases of E93 are also indicated in
Fig.~\ref{mmr}.  Selection becomes increasingly unlikely as one moves
from the dotted to the solid limits.

The result definitely leans towards the absence of low-metallicity,
low-age stars. Although the selection biases affect the region of
interest, the observed envelope clearly favours the low-dispersion AMR
model. The most significant evidence is the lack of metal-poor,
1.1-1.3 $M_\odot$ stars. This absence is best explained by the fact
that 1.1-1.3 $M_\odot$ stars, with maximum ages in the 5-8 Gyr range,
are too young to have experienced metallicities as low as $\feh=-0.6$.
Consequently, the objects observed in "Region II" in the AMR plot are
lower-mass objects, scattered from higher ages by the observational
error. A fact fully compatible with their age probability pdf.

This is a solid indication that the lack of definite AMR in the E93
sample is only apparent, independently of the discussion of the age
probability pdf. It should be confirmed with samples with wider
selection criteria and more objects, for instance by obtaining precise
spectroscopic metallicities for a sample of 1.1-1.3 $M_\odot$ stars
with a higher temperature cut.






\section{The dispersion in the AMR from Bayesian model comparison}
\label{model}

As the present study has indicated, the derivation of the AMR and its
intrinsic dispersion from the age-metallicity plot is made difficult
by the shape of the age uncertainties. Replacing prior-dependent age
estimates with large and asymmetrical uncertainty distributions by
single points makes direct "eyeball" analysis unreliable, and does not
permit the collection of the metallicity data into separate age bins.

However, this does not imply that the data cannot be used to study the
AMR. The posterior age pdf contains all the information available on
the ages, and there are other ways to analyse the data and constrain
the dispersion of the AMR, such as the mass-metallicity relation used
in Section~\ref{regions}.

The Bayesian framework also provides tools for the comparison of
different models. Let us call ${\cal M}$ the model assuming a
particular AMR with an intrinsic range $\Delta \feh$. The model
consists of
\begin{itemize}
 \item a mean age-metallicity relation: $\feh=f(t)$ with an intrinsic
  range $\Delta [Fe/H]$,
 \item stellar evolution models $L={\cal F}_L(m,t,z)$ and $T={\cal
    F}_T(m,t,z)$,
 \item assumptions about the distribution of the observational
  uncertainties (for instance Gaussian on $\log T$, $M_V$ and
  $[Fe/H]$).
\end{itemize}

What one wants to compute is $\Prob({\cal M} \mid D)$, the probability
of the model ${\cal M}$ being true, given all the data $D$. In
practice, one is not interested in the normalized probabilities but
wishes to compare the probabilities of two models.

Using Bayes' theorem,
\[
\Prob({\cal M} \mid  D)= \Prob({\cal M}) \frac{\Prob(D \mid {\cal M})}{\Prob(D)}
\]
the ratio of the probabilities for two models ${\cal M}_1$ and ${\cal
  M}_2$ is:

\[
\frac{\Prob({\cal M}_1 \mid  D)}{\Prob({\cal M}_2 \mid  D)}= \frac{\Prob({\cal M}_1)}{\Prob({\cal M}_2)} \frac{\Prob(D \mid {\cal M}_1)}{\Prob(D \mid {\cal M}_2)}
\]
where the unknown normalization $\Prob(D)$ vanishes in the ratio.

If the data points are independent, then the global term, ${\Prob(D
  \mid {\cal M})}$ can be broken down into a product of individual
probabilities for the individual data points $d_i$.
\[
{\Prob(D \mid {\cal M})}=  \prod_{i=1}^{N} \Prob (d_i |{\cal M})
\]
As in Section~\ref{compute} we marginalize over the mass and
metallicity, but now also over the age:
\[
\Prob(d_i \mid {\cal M}) = \int \!\!\! \int \!\!\! \int \Prob (d_i,m,t,z| {\cal M})\, \dd m\, \dd t\, \dd z
\]
Using the probability product rule:
\[
\Prob(d_i \mid {\cal M}) =\]\[
\int \!\!\! \int \!\!\! \int \Prob (d_i | {\cal M},m,t,z)\, \Prob(m,t,z | {\cal M})\, \dd m\, \dd t\, \dd z 
\label{bayesmodel}
\]
The first term in the integral is the likelihood, 
\[
 \Prob (d_i | {\cal M},m,t,z)\equiv {\cal L} (T_i,L_i,\feh_i, m,t,z) \]
and the second term is the prior in $(m,t,z)$ according to model
${\cal M}$, which we note $\rho(m,t,z)$.  Then,
\[
\Prob({\cal M} \mid  D) = \Prob({\cal M}) \]
$$\times \prod_{i=1}^{N}  \int\!\!\!\int\!\!\!\int \rho (m,t,z) {\cal L}(T_i,L_i,\feh_i, m,t,z)\, \dd m\, \dd t\, \dd z$$

As in Section~\ref{compute}, the integral over the $(m,t,z)$ space can
be evaluated by a Monte Carlo method:
\[
\Prob({\cal M} \mid  D) = \Prob({\cal M}) \prod_{i=1}^{N}  \sum_{j=1}^{n} {\cal L} (T_i,L_i,\feh_i,m_j,t_j,z_j)  
\]
where the $(m_j,t_j,z_j)$ triplets are $n$ draws according to the
probability distribution function $\rho(m,t,z)$.

The final step is to compute $\Prob({\cal M})$, the prior probability
of the model. As shown for instance by Sivia (1996), in the case of
varying the $\Delta \feh$ parameter, $\Prob({\cal M})$ is simply
inversely proportional to $\Delta \feh$. In the case of a
dispersionless relation with $\Delta \feh=0$, $\Prob({\cal M}) \sim
(\sqrt{2\pi}\sigma/\sqrt{N})^{-1}$, where $\sigma$ is the
observational uncertainty and $N$ the number of data points.

The Bayesian posterior probability $\Prob({\cal M}|D)$ was computed
for a set of models with different assumptions on the AMR, using the
same parameters as in Section~\ref{agese93}, with $10^6$ draws on the
Monte Carlo integration. The results are displayed in
Table~\ref{table1}. The probabilities are given relative to Model~1.
Model~1 assumes no relation between age and metallicity (flat AMR),
with a total metallicity range of $\Delta \feh$=0.80 dex. Model~2 is a
dispersionless AMR, linear in $z$, with $\feh=-0.3$ at $t=2.5$ Gyr and
$\feh =-0.8$ at $t=13$ Gyr. Model~3 assumes the same AMR, but with a
flat range of $\Delta \feh$=0.25 in the metallicities at a given age
(standard deviation $\sigma_{[Fe/H]}\simeq 0.07$ dex). Model 4 is the
same AMR, with $\Delta \feh$=0.40 ($\sigma_{[Fe/H]}\simeq 0.12$ dex).
Model~5 is an AMR of inverse slope, for comparison. To concentrate on
intermediate-age, thin disc objects -- the objects for which E93
indicate a high scatter -- we use the selection criteria $mass<1.2
M_\odot$ (removing very young objects) and $R_m>7$ kpc (removing thick
disc objects), where $R_m$ is the mean radius of the galactic orbit
computed by E93.

Table~\ref{table1} gives the logarithm of the resulting probabilities,
$\log_{10} P({\cal M} \mid D)$. The probabilities were computed
assuming $\sigma_{[Fe/H]}=0.075$, $\sigma_{log T}=0.009$, and
$\sigma_{M_V}$=0.15. Two objects have posterior probabilities
$10^{-3}$ below the maximum for model~3. Such outliers have an
excessive weight in the Bayesian model comparison because Gaussian
uncertainty distributions are assumed in the likelihood. In the real
world, unaccounted causes such as binarity or
misidentification leads to uncertainty distributions that have flatter
wings than Gaussians. The calculations were therefore also done
without these two objects (column "clipped" of Table~\ref{table1}).

The total probability was also computed allowing for the possible
presence of undetected binaries. As an upper limit to the possible
contamination, a proportion of 5 percent of equal-mass binaries was
assumed (fifth column of Table~\ref{table1}). Finally, the
calculations were also done with another set of assumed observational
uncertainties, $\sigma_{[Fe/H]}=0.10$, $\sigma_{log T}=0.01$, and
$\sigma_{M_V}$=0.10.

\begin{table*}
\begin{tabular}{l l l|r r r r}
Model &  $\Delta \feh$ & $\sigma_{[Fe/H]}$ &  ln$\,\Prob$(Model) & & & \\  
& & & raw & clipped & with 5\% & larger \\
& & &    & 2 objects & binaries & errors \\\hline \hline
Model 1 & 0.80 & 0.24 &     0.00  &    0.00  &    0.00  &    0.00  \\
Model 2 & 0    & 0    & $-$18.30  & $-$14.75 & $-$11.19 & $-$9.27  \\
Model 3 & 0.25 & 0.07 &  $-$2.94  &   +0.32  & $-$2.75  & $-$0.78  \\
Model 4 & 0.40 & 0.12 &  $-$0.63  &   +2.24  & $-$0.53  &   +1.06  \\
Model 5 & 0	& 0   & $-$46.36  & $-$40.45 & $-$41.93 & $-$29.20 \\\hline
\end{tabular}
\caption{
Total logarithmic posterior probability $\log_{10} \Prob({\cal M}|D)$,
for different models ${\cal M}$ of the age-metallicity relation, given
the data $D$ about the E93 sample.  Model 1: no significant AMR and
high intrinsic scatter; model 2: dispersionless linear AMR; models 3
and 4: linear AMR with low and medium intrinsic dispersion; model 5:
inverse dispersionless AMR. "Raw": all data with $mass<1.2 M_\odot$
and $R_m>7$. "Clipped": without HD~84737 and HD~88986.  "with 5\%
binaries": assuming 5 percent undetected equal-mass
binaries. "Other errors": assuming $\sigma_{[Fe/H]}=0.10$,
$\sigma_{log T}=0.01$, and $\sigma_{M_V}$=0.10
}

\label{table1}
\end{table*}

Table~\ref{table1} shows that the low-, medium- and high-dispersion
models are within one or two decades of each other in total
probability. The Bayesian computation shows that the model with a
range of 0.4 dex in metallicity at a given age, implying a standard
dispersion of 0.12 dex around a single AMR, is as favoured by the data
as the high-intrinsic scatter AMR within reasonable variations in the
assumptions.  Therefore, the data do not clearly favour a high
dispersion model over a low dispersion model of the AMR when the whole
age pdf's are taken into account.

\section{Conclusion} 

\subsection{Conclusions about E93 and the Galactic AMR}
\label{newe93}

The conclusion of our reappraisal of the implication of the E93 sample
for the age-metallicity relation in the Galactic disc is that the data
provide no solid evidence for the presence of a $\sigma\sim 0.24$
metallicity range at fixed age (or a $\Delta \sim 0.7$ range), as
usually stated. On the contrary, new data and a Bayesian age
determination put an upper limit of $\sigma=0.15$ dex on the intrinsic
scatter of the AMR. An extended Bayesian probability analysis shows
that the age probability distributions are much wider than realized,
and that visual interpretation of an age-metallicity plot like
Fig.~\ref{amr_e93} is likely to be misleading. The age uncertainties
are also too large for an age binning of the data to be made with any
confidence.

New Hipparcos parallax and Coravel radial velocity data on the same
sample confirm the doubts introduced by the Bayesian approach, and
show that many outliers on the age-metallicity diagram are indeed
detected binaries or stars with either discrepant distance estimates
or discrepant metallicity estimates. Many ages are also put nearer to
the mean AMR by the temperature adjustments found necessary between
the stellar evolution models and the observations. This, together with
a Bayesian model-testing analysis, point to a rather well-defined AMR
with a smaller metallicity gradient at fixed age, with a standard
deviation of the order of 0.15 dex or lower, or a total range of $
<\sim$0.4 dex at a given age.

This lower range is confirmed by examining the behaviour of the data
in two specific zones of the mass-metallicity plot, showing the
absence of young, metal-poor stars in the $\feh \sim -0.5$, $t\leq 6$
Gyr zone, and of old, solar-metallicity stars in the $\feh \sim 0.0$,
$t \geq 5$ zone.

The implication is that there is no mandatory need at this point for
galactic models to reproduce a very large scatter of metallicity in
the ISM at a given time and galactocentric radius for the Galactic
disc.  It restores the coherence with the numerous other indications
of a low present-day dispersion in the abundance of the gas (e.g. ISM,
cepheids, HII regions, see Introduction for references). The remaining
dispersion is still quite large, and shows that a simple, single-AMR
model is not sufficient. But it lies within the values observed in
other star-forming galaxies, and indicated by Galactic open clusters.
It is also within the scale of what reasonable chemical
inhomogeneities and radial orbital mixing can achieve without the need
to invoke long-lived extreme inhomogeneities or infall in the past.

\subsection{General Conclusions and recommendations}
\label{recom}

Looking beyond the E93 sample to future studies of the chemical and
dynamical history of the Galactic disc, we now consider some
implications of our results.

Metallicities with an internal uncertainty of $\sim 0.05$ dex, as in
E93, with Hipparcos distances ($\sigma_\mu\sim 0.1$ mag at 50 pc) are
still about as accurate as can presently be achieved in terms of
uncertainties of the observables.  Colour-temperature transformations,
bolometric corrections and model temperature errors are also sources
of uncertainties that are proving difficult to reduce below the level
of 0.01 dex on $\log T$ and 0.10 mag on $M_V$.

With these kinds of accuracies, the posterior age pdf's are often wide
and asymmetrical, especially for later-type stars -- the ones most
useful in the study of the history of the Galaxy. In that case, ages
computed with the standard method can be strongly biased, and
replacing the full probability distribution by a single central value
can lead to misleading impressions.

For large samples, uncertainties of about 0.10 dex or larger in [Fe/H]
are more typical (see for instance Ibukiyama \& Arimoto 2002, Feltzing
et al. 2001), implying even wider age pdf's.  In this case, it should
be realized that when the probability distributions for the ages are
much wider than the dispersion of the points themselves, adding more
points only provides a better definition of these probability
distributions themselves, without actually adding much information on
the underlying age distribution.  This regime dominated by systematic
effects is clearly apparent in fig.~5 of Ibukiyama \& Arimoto (2002)
and fig.~10 of Feltzing et al. (2001) as the "wave-shape" in the
age-metallicity diagram. Not only is a mean metallicity {\it
  decreasing} with time near 5-10 Gyr difficult to understand in terms
of galactic evolution, but it is also exactly the kind of shape that
we expect with a bias towards the terminal age (the "Region I" in
Fig.~\ref{amr_e93}). Such a revealing shape is also apparent in the
AMR plot of \GC. Thus, as correctly reckoned by E93,
a small, low-error sample is preferable in this regime to a large,
high-error sample.

We also note that selections of subsamples by imposing a limit on the
relative age error, e.g. $\sigma_{\mtage}/\mage < 0.5$ as in Feltzing
et al. (2001), should be avoided, because they strongly reinforce the
"terminal age bias" (see Section~\ref{laurent}). Because
$age_{\mbox{\tiny true}}$ is not accessible, the selection is in fact
$\sigma_{\mtage}/\mage_{\obs} < limit$, which favours ages near the
upper limit of their error bar with a low {\it apparent}
$\sigma_{\mtage}$. For instance, in our Fig.~\ref{fig:termage}, such a
selection would pick up only the most strongly biased ages with
$\mage_{\obs}\sim 10$ Gyr.

The following suggestions are proposed for the computation of
isochrone ages and the study of the history of the Galactic disc:

\begin{description}
\item -- For late-type stars, the posterior age pdf can be computed
  rather than the "nearest isochrone" age which can be strongly
  biased.
\item -- Smaller samples with lower uncertainties should be preferred to
  large samples with higher uncertainties.
\item -- The full age pdf should replace Gaussian approximations to
  examine the compatibility of the data with a given hypothesis. The
  age pdf often has wide and flat wings. With such strongly ungaussian
  distributions, mathematical hypothesis-testing can be sounder than
  eyeball analysis.
\item -- The mass vs. age plot can be used as a diagnostic. If the
  derived ages cluster towards the end-of-main-sequence lifetime, they
  are probably subject to a strong systematic bias ("terminal age
  bias").
\item -- Relative error selection criteria ($\sigma_{\mtage}/\mage <
  limit$) should be avoided to form subsamples with better determined
  ages.  $\sigma_{\mtage} < age_{\mbox{\tiny MS}}$, where
  $\mage_{\mbox{\tiny MS}}$ is the main sequence lifetime at the
  star's mass, is a good alternative.
\end{description}

Our results also tend to rehabilitate the method of age determination
from chromospheric activity. Discrepancy between chromospheric and
isochrone ages had led to some suspicion of unrecognized uncertainties
in the former method (see Introduction). However, according to our
study, a large part of the mismatch can be attributed to the
systematic effects affecting direct isochrone ages. The
age-metallicity relation using chromospheric ages shows a lower
scatter, adding further confidence in the reliability of chromospheric
ages. A detailed Bayesian comparison of isochrone and chromospheric
ages would be useful in this context.

Finally, our study suggests that given the high sensitivity of the age
determination to observational uncertainties and in particular to
statistical outliers, it can be very useful to combine independent
determinations of the input quantities - temperature, luminosity,
metallicity - in order to attempt to identify the objects which may be
such outliers. The strongest biasing effects are highly non-linear and
can be much reduced by removing such objects.

An interactive code to compute Bayesian age estimates for Galactic
dwarfs is available at the following website:
\begin{verbatim}
http://obswww.unige.ch/~pont/ages
\end{verbatim}

\section{Acknowledgements}
We would like to thank Antonio Aparicio for providing the stellar
evolution synthesis code IAC-star in advance of publication.
F.P. is indebted to the Instituto de Astrof\'\i sica de Canarias,
for a scientific stay in 2003, and in particular to Carme Gallart
 for making this stay possible and enjoyable.

\end{document}